\documentclass[twocolumn,trackchanges]{aastex631}

\newcommand{\mytilde}{\raise.19ex\hbox{$\scriptstyle\sim$}}
\newcommand{\mcc}{MC$^{2}$}

\newcommand{\kms}{km~s$^{-1}$}
\newcommand{\h}{$h_{70}^{-1}$}
\newcommand{\Msun}{$M_{\sun}$}

\usepackage{amsmath}

\accepted{}

\shorttitle{Multiwavelength Analysis of A1240}
\shortauthors{Cho et al.}

\begin{document}

\title{Multiwavelength Analysis of A1240, the Double Radio Relic Merging Galaxy Cluster Embedded in a $\mathbf{\mytilde}$80~Mpc-long Cosmic Filament}

\correspondingauthor{M. James Jee}
\email{hyejeon@yonsei.ac.kr, mkjee@yonsei.ac.kr}

\author[0000-0001-5966-5072]{Hyejeon Cho} 
\affiliation{Department of Astronomy, Yonsei University, 50 Yonsei-ro, Seodaemun-gu, Seoul 03722, Korea}

\author[0000-0002-5751-3697]{M. James Jee} 
\affiliation{Department of Astronomy, Yonsei University, 50 Yonsei-ro, Seodaemun-gu, Seoul 03722, Korea}
\affiliation{Department of Physics and Astronomy, University of California, Davis, One Shields Avenue, Davis, CA 95616, USA}

\author[0000-0001-5303-6830]{Rory Smith}
\affiliation{Korea Astronomy and Space Science Institute (KASI), 776 Daedeokdae-ro, Yuseong-gu, Daejeon 34055, Korea}
\affiliation{University of Science and Technology (UST), 217 Gajeong-ro, Yuseong-gu, Daejeon 34113, Korea}

\author[0000-0002-4462-0709]{Kyle Finner}
\affiliation{Infrared Processing and Analysis Center, California Institute of Technology, Pasadena, CA 91125, USA}
\affiliation{Department of Astronomy, Yonsei University, 50 Yonsei-ro, Seodaemun-gu, Seoul 03722, Korea}

\author[0000-0002-1566-5094]{Wonki Lee}
\affiliation{Department of Astronomy, Yonsei University, 50 Yonsei-ro, Seodaemun-gu, Seoul 03722, Korea}


\begin{abstract}
We present a multiwavelength study of the double radio relic cluster A1240 at $z=0.195$. Our Subaru-based weak lensing analysis detects three mass clumps forming a $\mytilde4$~Mpc filamentary structure elongated in the north-south orientation. The northern ($M_{200}=2.61_{-0.60}^{+0.51}\times10^{14} M_{\sun}$) and middle ($M_{200}=1.09_{-0.43}^{+0.34}\times10^{14} M_{\sun}$) mass clumps separated by \mytilde1.3~Mpc are associated with A1240 and co-located with the X-ray peaks and cluster galaxy overdensities revealed by $Chandra$ and MMT/Hectospec observations, respectively. The southern mass clump ($M_{200}=1.78_{-0.55}^{+0.44}\times10^{14} M_{\sun}$), \mytilde1.5~Mpc to the south of the middle clump, coincides with the galaxy overdensity in A1237, the A1240 companion cluster at $z=0.194$. Considering the positions, orientations, and polarization fractions of the double radio relics measured by the LOFAR study, we suggest that A1240 is a post-merger binary system in the returning phase with the time-since-collision $\mytilde1.7$~Gyr. With the SDSS DR16 data analysis, we also find that A1240 is embedded in the much larger-scale ($\mytilde80$~Mpc) filamentary structure whose orientation is in remarkable agreement with the hypothesized merger axis of A1240.
\end{abstract}

\keywords{Abell clusters (9); Cosmic web (330); Dark matter (353); Intracluster medium (858); Radio continuum emission (1340); Weak gravitational lensing (1797); X-ray astronomy (1810)}

\defcitealias{Barrena+2009}{B09}
\defcitealias{Golovich+2019b}{G19b}
\defcitealias{WH15}{WH15}

\section{Introduction} \label{sec:intro}

Radio relics, also known as cluster radio shocks, refer to Mpc-scale diffuse synchrotron radio emission features observed in the outskirts of merging galaxy clusters. 
Tracing the merger shocks, they provide unique information, such as viewing angle, collision velocity, merger axis, etc., on the merging scenarios that cannot be directly accessed with other wavelength data alone (e.g., \citealt{Ensslin+1998}; see \citealt{vanWeeren2019review} for a recent review and references therein). There is a growing interest in studying radio relic clusters to understand the formation and evolution of the merger shocks, the mechanism of cosmic ray particle acceleration, the interaction between plasma turbulence and intracluster magnetic field, etc. In addition, to those who desire to utilize merging clusters as dark matter laboratories, radio relic clusters provide powerful environments because the information on their merger histories is more readily available \citep[e.g.,][]{Ng+2015,Golovich+2016MC2_MACSJ1149,Monteiro-Oliveira+2017_A3376,Finner+2021}.

Double radio relics are a subclass of radio relics that exhibit two relics on the opposite sides. 
The first detection of double relics dates back to 1997, when \citet{Rottgering+1997_A3667} found two extended diffuse radio sources in the northwest and southeast peripheral regions of A3667 using ATCA and MOST data.
Although in principle every collision between clusters generates shocks in pairs, observationally there are only a dozen or so double radio relic systems reported to date. The rarity indicates that the merger perhaps should meet some delicate criteria in order to produce two observable radio relics.

Given that our understanding of the exact mechanism in relic creation and particle acceleration is still highly incomplete, the double radio relic system serves as a powerful testbed for different models because the two relics probe the same merger. In particular, {\it symmetric}\footnote{Here, we define symmetric radio relics as the system where the two normal vectors to the radio relics agree and are consistent with the hypothesized merger axis.} double radio relic systems are most useful, as they allow us to reduce the complexity in our inference of their merger scenarios. Three well-known examples in this category are PSZ1~G108.18-11.53, MACS~1752.0+4440, and ZWCL~1856.8+6616 \citep[e.g.,][]{deGasperin+2015_PSZ1G108.18-11.53,Finner+2021}. 

\begin{figure*}
\begin{center}
\includegraphics[width=0.75\linewidth]{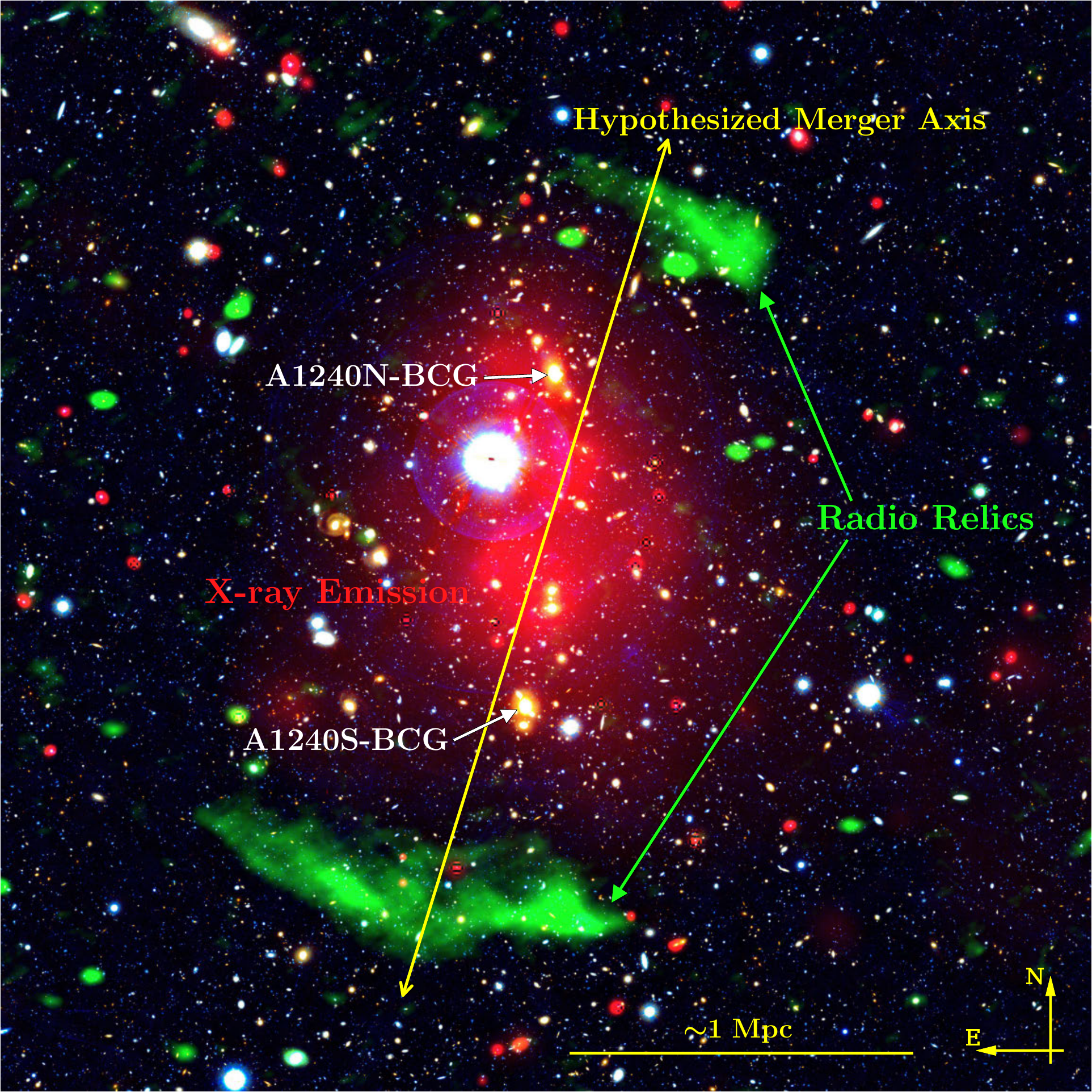}
\caption{LOFAR, Subaru, and {\it Chandra} view of the merging cluster A1240.
The composite image shows the $16\farcm5\times16\farcm5$ (3.2~Mpc $\times$ 3.2~Mpc) region centered on A1240. The intensity in green represents the 143~MHz radio emission measured with LOFAR \citep{Hoang+2018}. The intensity in red represents the {52~ks} {\it Chandra} X-ray surface brightness map.
The background RGB image is created using the Subaru/Suprime-Cam $r$, $r+g$, and $g$ filter data for the red, green, and blue channels, respectively. The hypothesized merger axis is aligned with the X-ray elongation and also perpendicular to both radio relics.
\label{fig:a1240}}
\end{center}
\end{figure*}

In this paper, we present multi-wavelength analysis of the cluster merger A1240 at $z=0.195$, which belongs to the Merging Cluster Collaboration (\mcc) {\it gold sample} that the collaboration decided to follow up with priority by more detailed studies including comparisons between their galaxy and mass distributions \citep[][hereafter G19b]{Golovich+2019b}. 
A1240 is one of the few systems that possess symmetric double relics (Figure~\ref{fig:a1240}). 

The first suggestion that A1240 might host two radio relics was made by \citet{Kempner+Sarazin2001} based on the WENSS and NVSS radio images. 
With deep VLA observations at 20~cm (1.4~GHz) and 90~cm (325~MHz), \citet{Bonafede+2009} confirmed the presence of the two radio relics in A1240 and firmly detected a spectral index flattening towards the cluster outskirt in the northern relic, which is consistent with our expectation if the relic indeed traces the merger shock travelling to the north. 
For both relics, moderate Mach numbers of \mytilde3 were derived. \citet{Bonafede+2009} also found that the mean level of fractional polarization in the relics is high (\mytilde30\%), reaching up to 70\%; in general, a smaller angle between the merger axis and the plane of the sky is believed to produce a larger polarization fraction.
\citet{Hoang+2018} used high-resolution and wide-frequency range (LOFAR 143~MHz, GMRT 612~MHz to VLA 3~GHz) observations and found that the spectral indices of {\it both} relics steepen from the outer edges towards the cluster center. 
\citet{Hoang+2018} revealed that the projected largest linear size and radio power of the southern relic are twice larger and three times higher than those of the northern relic, respectively. 
From the measurement of the mean fractional polarization, they estimated the viewing angle\footnote{In this study, the viewing angle $\alpha$ is defined to be the angle between the merger axis and the plane of the sky. Note that the definition used by \citet{Hoang+2018} is $90\degr-\alpha$.}
to be $\alpha\lesssim51\degr$.

Now, to complete the merging scenario of A1240, among the most important missing data are its dark matter (DM) substructures and masses. The mass is critical in estimating the time scale of the merger because it determines the collision speed, which is related to the shock propagation speed and the position of the relics.
As the merging system departs from the hydrostatic/dynamic equilibrium, mass estimates based on X-ray, Sunyaev-Zel'dovich (SZ), or galaxy-velocity observations are expected to produce biased results. To address the issue, we perform weak-lensing (WL) analysis with Subaru/Suprime-Cam imaging data. 
No measurement of the WL signal has been reported to date for the A1240 field. Our WL study of A1240 and its subclusters provides independent mass estimates.
Although WL mass estimates are not totally bias-free, the method is least affected by the dynamical state of the system.

In addition to WL, we investigate the cluster galaxy position/velocity distribution of A1240 with our MMT/Hectospec data. \citet[][hereafter B09]{Barrena+2009} used the TNG-DOLORES/MOS and SDSS DR7 data and identified the north-south bimodal structure with 62 cluster members in A1240. Furthermore, 27 members were assigned to its companion cluster A1237.
\citetalias{Barrena+2009} estimated the line-of-sight (LOS) velocity difference between the two substructures to be $\mytilde400$~\kms. 
\citet{Golovich+2019a,Golovich+2019b} increased the total number of the spectroscopic members of A1240 to 146 using Keck/DEIMOS observations, confirming both the bimodality and LOS velocity reported by \citetalias{Barrena+2009}. \citetalias{Golovich+2019b} also identified 24 member galaxies for A1237. 
Our MMT/Hectospec observation brings up the total confirmed members to $\mytilde432$ in the A1240 field (A1240$+$A1237), which is a significant (a factor of 2.5) increase. Aided with our Subaru imaging data, we refine our understanding of the cluster substructure and dynamical state.
Using the SDSS spectroscopic catalog of galaxies, we also explore the surrounding structures of A1240 out to the clustercentric radius of \mytilde100~Mpc.
Finally, combining our multi-wavelength observations from radio to X-ray with Monte-Carlo analysis, we constrain the merging scenario of A1240. 

Throughout the paper, we adopt a flat $\Lambda$CDM cosmology with $H_{0}=70$~\kms~Mpc$^{-1}$, $\Omega_{M}=0.3$, and $\Omega_{\Lambda}=0.7$. 
At the redshift of A1240, $z=0.195$, an angular scale of 1\arcsec\ corresponds to a physical scale of \mytilde3.24~kpc (\mytilde194~kpc~arcmin$^{-1}$). 
Unless otherwise specified, north is up and east is left in our two-dimensional image presentations. All magnitudes are in the AB magnitude system. 
The $M_{200}\ (M_{500})$ value refers to a halo mass within a spherical volume of radius $r_{200}\ (r_{500})$, at which the mean cluster density is 200 (500) times the critical density of the universe at the redshift of the cluster. 

We structure our paper as follows. \textsection\ref{sec:obsdata} describes our data and reduction methods. We present our analysis of the cluster member distributions and WL signal in \textsection\ref{sec:analysis}. Our results are compared with previous studies and we present a new merging scenario in \textsection\ref{sec:discussion} before we conclude in \textsection\ref{sec:summary}.

\section{Observations} \label{sec:obsdata}

\subsection{Subaru/Suprime-Cam} \label{subsec:sscobs}

The A1240 field was observed with Subaru Prime Focus Camera \citep[Suprime-Cam;][]{SuprimeCam2002} on 2014 February 25 in $g$- and $r$-bands (PI: D. Wittman) with total integrations of 720~s and 2880~s, respectively, as part of the \mcc\ radio-relic-selected optical imaging survey \citep{Golovich+2019a}. 
We used dithering and field rotation among four and eight pointings for $g$ and $r$, respectively. This not only reduces the effects of cosmic rays, but also distributes the saturation trails and diffraction patterns from bright stars azimuthally, which are removed in the later stacking process. 
The reliability of this scheme for detecting faint, small galaxies near bright stars, 
has been demonstrated in a series of our Subaru/Suprime-Cam WL studies 
\citep[e.g.,][]{Jee+2015, Jee+2016, Finner+2017, Finner+2021}. 

We used the SDFRED2 package\footnote{\url{https://www.naoj.org/Observing/Instruments/SCam/sdfred/sdfred2.html.en}} \citep{sdfred2} for the low-level data reduction steps such as overscan and bias subtraction, flat fielding, geometric distortion correction, atmospheric dispersion correction, auto-guider shade masking, etc. 
We obtained astrometric solutions for individual frames 
using SExtractor\footnote{\url{https://www.astromatic.net/software/sextractor}} \citep{sextractor} and 
SCAMP\footnote{\url{http://www.astromatic.net/software/scamp}} \citep{scamp} with Sloan Digital Sky Survey (SDSS) DR9 as a reference. 
To create final mosaic images, we performed a two-step process of median-stacking and weighted-averaging co-addition for each filter with the SWarp software\footnote{\url{http://www.astromatic.net/software/swarp}} \citep{swarp}. 
See our previous \mcc\ studies \citep[e.g.,][]{Jee+2015, Jee+2016, Finner+2017} for details about Suprime-Cam data reduction procedures. 

For both $g$- and $r$-band mosaic images, we performed object detection and photometry with SExtractor in dual-image mode.
We used the deeper $r$-band image and its associated output weight-map from SWarp for object detection.
The dual mode ensures that the same object positions and apertures are used for the photometry in different filter sets. We identified objects with an area of at least five connected pixels above two times the local background rms. 
Separation of blended objects was performed using {\tt DEBLEND\_NTHRESH}=64 and {\tt DEBLEND\_MINCONT}=$10^{-4}$.
For the photometric measurements, we used the rms maps, which were 
constructed from the SWarp-produced weight maps. 

Since Subaru/Suprime-Cam filter throughputs are similar to those of SDSS, we calibrated the Subaru photometry using the star catalog from SDSS DR16, which covers the A1240 field. 
Since the SDSS magnitudes were corrected for Galactic extinction \citep{SF11}, our calibration also includes the correction.
We employed {\tt MAG\_AUTO} to estimate the total flux for each object. 
For the color estimation, we used the magnitude measured within the isophotal area ({\tt MAG\_ISO}). 
The limiting magnitudes where the detection completeness fraction reaches 50\% \citep{lmag5Harris} are 26.2~mag and 27.0~mag in {\tt MAG\_AUTO} for the $g$ and $r$ filters, respectively.

\subsection{MMT/Hectospec} \label{subsec:mmtobs}

Spectroscopic observations (PI: K. Finner) of the A1240 field were carried out with the 6.5-m MMT/Hectospec as part of the Korean GMT (K-GMT) project. The 300 fiber Hectospec \citep{Fabricant2005} instrument was configured for two pointings utilizing the 270 groove/mm grating with a spectral coverage of 3650--9200~\AA. Cluster member candidates within the one degree field of view of the instrument were selected from Subaru imaging, archival SDSS imaging, and SDSS photometric redshifts. For the Subaru and SDSS imaging, candidates were selected within $\Delta(g-r)\pm0.1$ of a linear fit to the existing spectroscopic redshifts in a $g-r$ vs. $r$ color--magnitude diagram. Photometric redshift candidates were selected within $z=0.195\pm0.05$. The candidates were split into a bright configuration with objects $r\leqslant20$ and a faint configuration with $r\leqslant21$. 
The bright configuration were observed on 2019 March 5 for one hour of integration and the faint configuration was observed on 2019 March 26 with two hours of integration, both on a twenty minute cadence.

We reduced the MMT/Hectospec observations with HSRED v2.1 pipeline\footnote{\url{https://github.com/MMTObservatory/hsred}}
provided by the Telescope Data Center (TDC) at SAO, originally written by Richard Cool. 
The Krypton lamp light, one of MMT PenRay emission calibration lamps, has affected part of observations as emission or absorption, and makes it difficult to estimate redshifts correctly. 
For this observation, we used the TDC-reduced data. 
We derived redshifts for each of 477 spectra through a cross-correlation routine with a set of template spectra using the {\tt RVSAO} add-on package in IRAF \citep{Kurtz1998}.

\begin{figure}
\begin{center}
\includegraphics[width=1.0\linewidth]{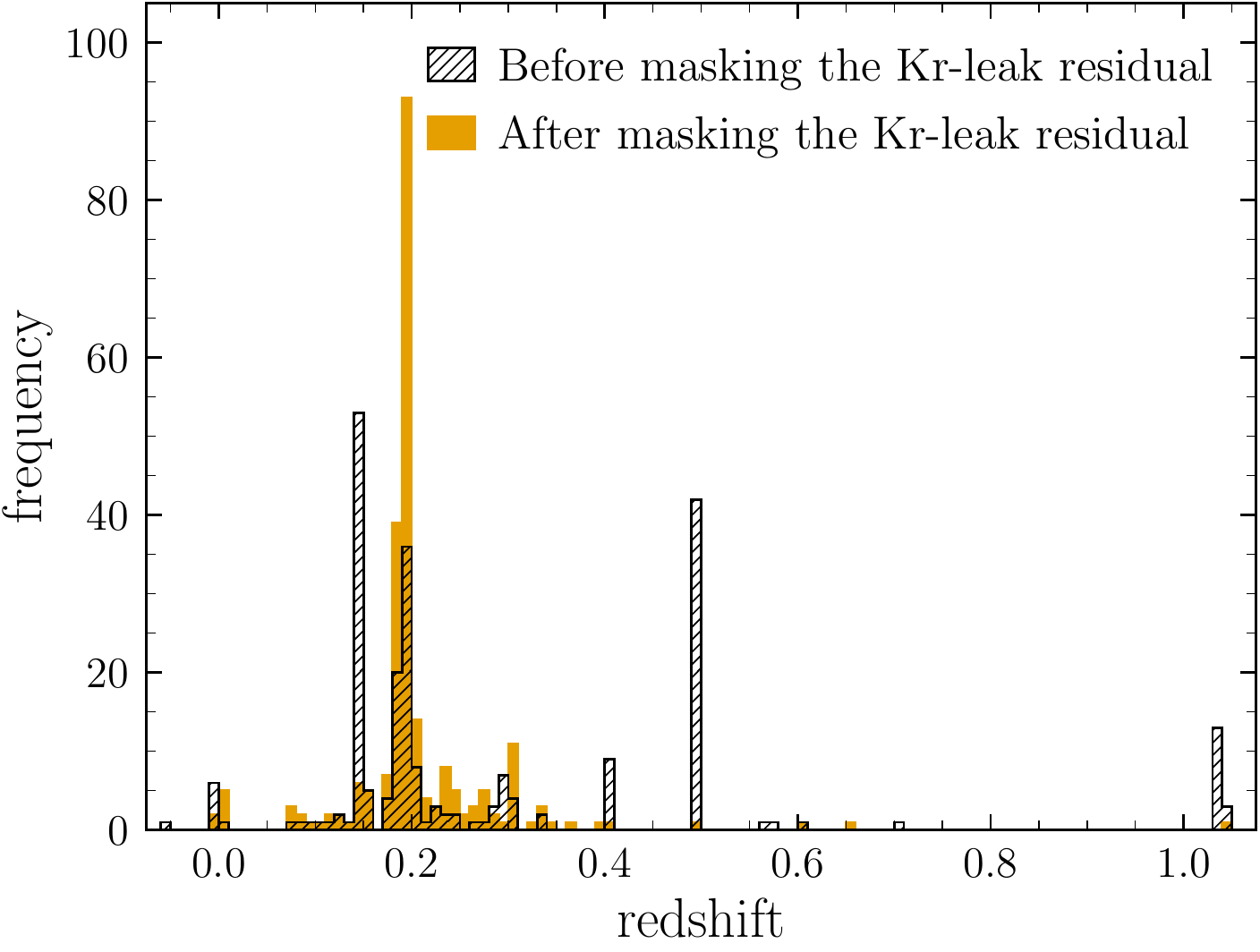}
\caption{Redshift distribution from the first-night MMT observation, which was affected by the Krypton lamp light. The hatched histogram is obtained from the redshift estimates before the calibration lamp light leak residuals were masked out. The filled orange is the result when we masked out the residuals. 
The number of cluster member candidates in the first-night observation increases from 64 (\mytilde$27\%$) to 149 (\mytilde$62\%$) after the correction for the light leaks was performed.
\label{fig:Kr-mask}}
\end{center}
\end{figure}

Figure~\ref{fig:Kr-mask} shows the effect of the Krypton light leak on the redshift measurements. 
Because some residual from the Krypton light leak is still evident in the TDC-cleaned spectra (hatched histogram), we provided a list of masked regions for the residual to estimate redshifts more reliably (orange histogram). 
After the residuals are masked out, 
the number of the A1240 cluster galaxy candidates increases from 64 (\mytilde$27\%$) to 149 (\mytilde$62\%$) in the bright configuration.

We adopted the cross-correlation score $r$-value \citep{Tonry1979} $R_{\rm XC}$, a measure of significance of the cross-correlation peak, greater than 4 as secure redshift measurements.
When we used the criteria of $R_{\rm XC}>4$ and $z_{\rm err}<0.001$ ($v_{\rm err}<300$~\kms),
we obtained 443 reliable redshift measurements from our new MMT observations including 236 member candidates.
In the following section, we describe the details of the membership determination. 

\section{Analysis} \label{sec:analysis}

\subsection{Cluster Galaxy Distributions} \label{subsec:galanal}

\subsubsection{Membership Determination} \label{subsubsec:membership}

To maximize the number of spectroscopically confirmed cluster galaxies, we collected publicly available spectroscopic redshifts in the A1240 field. 
\citetalias{Barrena+2009} provides redshifts of 145 galaxies. 
They combined 118 redshifts acquired at the TNG telescope and 32 publicly available SDSS galaxies (5 galaxies are in common). 
As part of the \mcc\ radio-relic-selected spectroscopic survey \citep{Golovich+2019a}, we observed A1240 with the DEIMOS multi-object spectrograph with the Keck II telescope on 2013 December 3 and 2015 February 16 (PI: W. Dawson). The spectroscopic observations were conducted with two slit masks for the SDSS-targeted galaxies. We obtained 188 spectroscopic redshifts of galaxies. 
In addition, from SDSS DR16 we retrieved 168 spectroscopic galaxies within $R = 25\arcmin$ of the position at ${\rm RA} = 11^{\rm h}23^{\rm m}36\fs0$, ${\rm Dec} = +43\arcdeg05\arcmin04\farcs2$.
We found that there are 37 galaxies in common among the different catalogs and compiled a total of 464 published redshifts.

\begin{figure} 
\begin{center}
\includegraphics[width=0.98\linewidth]{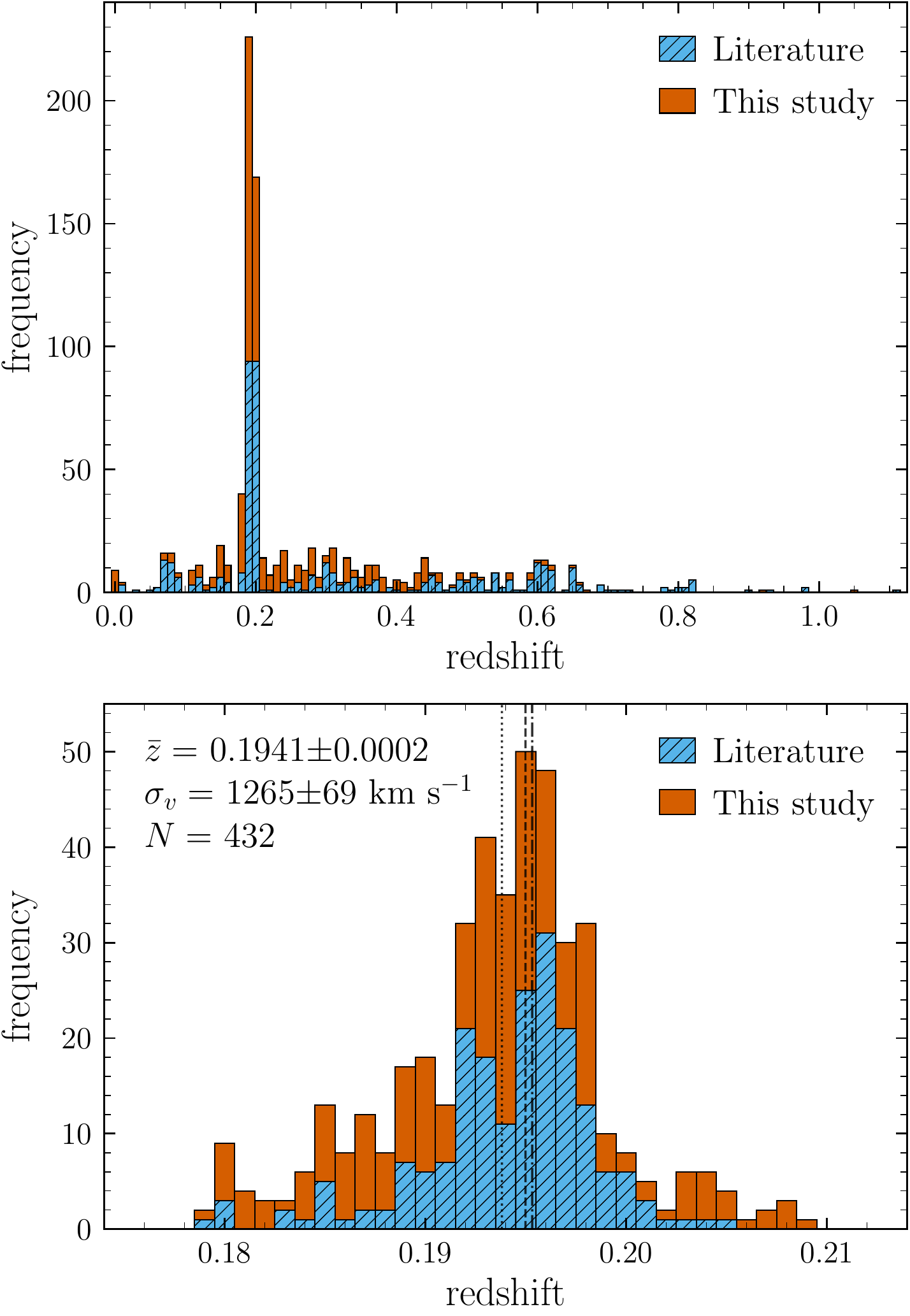}
\caption{Redshift distributions for our combined spectroscopic catalog in the A1240 field. 
The blue-hatched bars represent the redshift distribution of galaxies from the literature, including SDSS DR16 galaxies, while the orange-filled stacked bars are from our MMT/Hectospec observations. 
Duplicates are removed. 
The top panel shows all galaxies with spectroscopic redshifts listed in Table~\ref{tab:speccat}.
The redshift distribution of the member (candidate) galaxies is illustrated in the bottom panel. Spectroscopically confirmed members are determined by performing an iterative $3\sigma_{v}$ clipping process with the bi-weight estimator. 
The total number of member candidates is 432 with a global (bi-weight) mean redshift of $0.1941\pm0.0002$ and a rest-frame velocity dispersion of $1265\pm69$~\kms, along with $1\sigma$ uncertainties derived from 10000 bootstrap resamplings.
Bins have widths of $\Delta z = 0.01$ ({\it top}) and $\Delta z=0.001$ ({\it bottom}).
The dot-dashed and dashed vertical lines demonstrate robust mean redshifts of northern and southern luminosity-subclumps of A1240, whereas the dotted line shows that of A1237 luminosity clump (see Figure~\ref{fig:speczdist3sig}). 
}\label{fig:speczhist}
\end{center}
\end{figure}

We combined these spectroscopic redshifts from the literature and our MMT measurements in the A1240 field and created a final catalog of 934 galaxies. 
Out of the 464 published spectroscopic redshifts, 7 galaxies are also present 
in our MMT/Hectospec catalog and their values agree well up to the fourth decimal places ($\Delta z=0.000191$; $\Delta v=57.2$~\kms). 
We used our MMT redshift measurements and recently published spectroscopic redshifts for duplicated targets.
The distributions of galaxies in the combined spectroscopic catalog as a function of redshift are displayed in the top panel of Figure~\ref{fig:speczhist}. 
Table~\ref{tab:speccat} lists the redshift information of the final catalog.

\begin{table*} 
\caption{Spectroscopic Redshift Catalog of the A1240 Field\label{tab:speccat}}
\begin{tabular*}{\textwidth}{@{\extracolsep{\fill}}ccRcCrc}
\hline\hline
\multicolumn1c{R.A. (J2000)}\vspace{-0.35cm} & \multicolumn1c{Decl. (J2000)} & \multicolumn1c{$cz$} & & & & \multicolumn1c{Catalog} \\
 & & \multicolumn1c{} & \colhead{$z$}\vspace{-0.35cm} & \colhead{$z_{\rm err}$} & \colhead{$R_{\rm XC}$} & \\
\multicolumn1c{(deg)} & \multicolumn1c{(deg)} & \multicolumn1c{(\kms)} & & & & \multicolumn1c{Source} \\
\hline
170.2139000 & 42.9799575 & 80183\pm82 & 0.267460 & 2.72518\times10^{-4} & 3.1 & MMT \\
170.2442542 & 43.2169419 & 68205\pm33 & 0.227509 & 1.10282\times10^{-4} & 12.3 & MMT \\
170.2460167 & 43.1449014 & 34216\pm72 & 0.114131 & 2.38641\times10^{-4} & 3.9 & MMT \\
170.2488167 & 43.2495383 & 66035\pm11 & 0.220270 & 3.80040\times10^{-5} & 11.7 & MMT \\
170.2498042 & 42.8980903 & 131726\pm22 & 0.439392 & 7.40362\times10^{-5} & 11.3 & MMT \\
170.2548250 & 43.1294517 & 98990\pm13 & 0.330195 & 4.22305\times10^{-5} & 11.6 & MMT \\
170.2631375 & 42.9662894 & 58725\pm42 & 0.195887 & 1.40097\times10^{-4} & 7.6 & MMT \\
170.2673167 & 43.0931739 & 132779\pm28 & 0.442902 & 9.38973\times10^{-5} & 8.5 & MMT \\
170.2721083 & 42.8627586 & 59767\pm53 & 0.199360 & 1.75617\times10^{-4} & 7.9 & MMT \\
170.2747250 & 43.2392617 & 54690\pm21 & 0.182426 & 6.94714\times10^{-5} & 19.1 & MMT \\
\hline
\end{tabular*}
\tablecomments{Table~\ref{tab:speccat} is published in its entirety in the machine-readable format. A portion is shown here for guidance regarding its form and content. The columns list: (1) right ascension and (2) declination in degrees (J2000.0); (3) the redshift velocity and its measurement uncertainty in \kms; (4) the redshift and (5) its uncertainty; and (6) the cross-correlation score $r$-value obtained from {\tt RVSAO} (valid only for MMT catalog). The last column gives the references of spectroscopic data. The code MMT, B09, G19, and SDSS refer our MMT/Hectospec observations, \citetalias{Barrena+2009}, \citet{Golovich+2019a}, and SDSS DR16. 
}
\end{table*}

To separate cluster member candidates from foreground and background objects, 
we performed iterative 3-$\sigma$ clipping, estimating the mean redshift and velocity dispersion using the bi-weight estimator \citep{Beers+1990}.
In combination with the publicly available data, we obtained 432 member (candidate) galaxies in the A1240 field (stacked histograms in the bottom panel of Figure~\ref{fig:speczhist}), out of 898 reliable redshift measurements\footnote{The criteria are $R_{\rm XC} > 4$ and $v_{\rm err}<300$~\kms.}.

\subsubsection{Galaxy Number and Luminosity Distributions} \label{subsubsec:galdens}

\begin{figure}
\begin{center}
\includegraphics[width=1\linewidth]{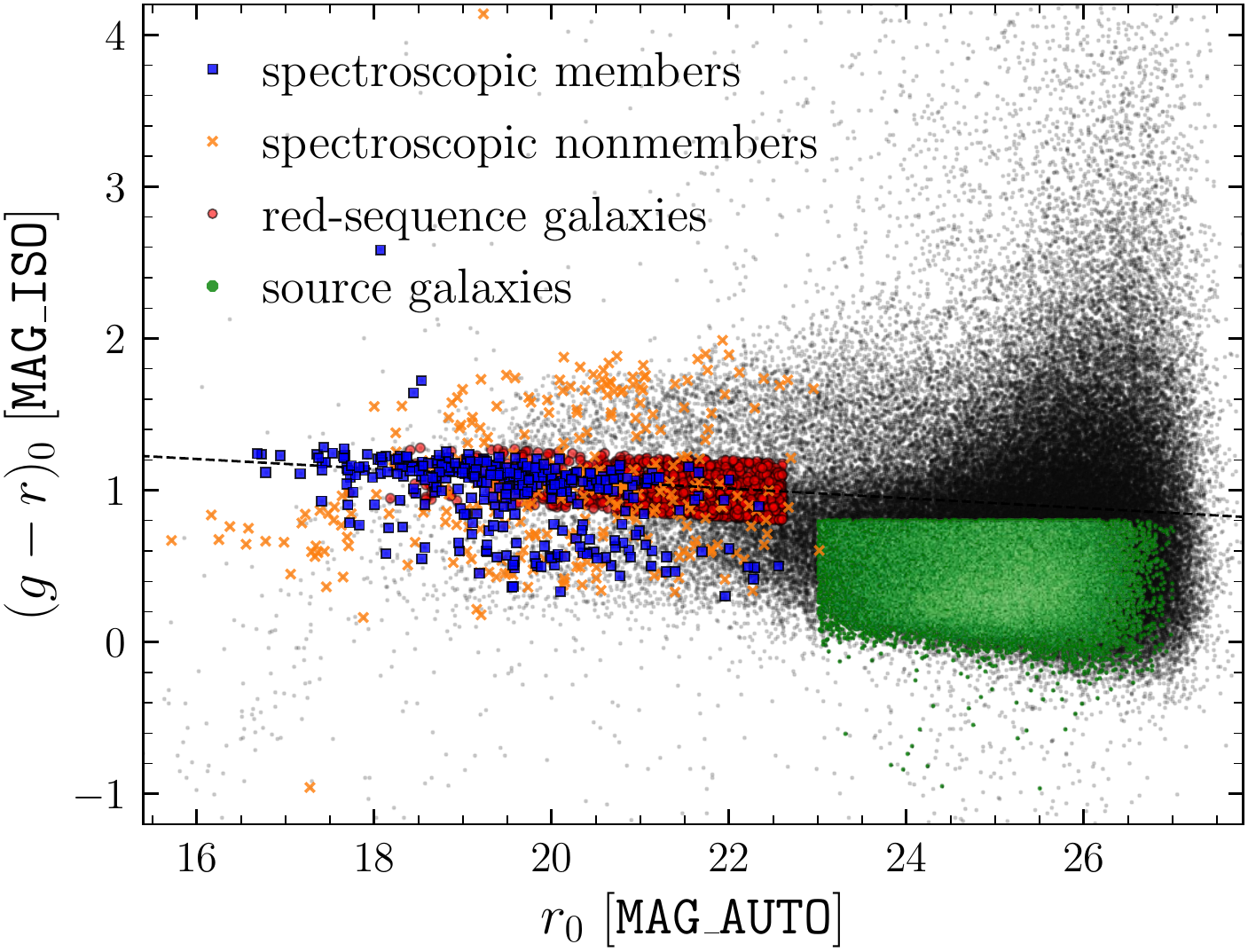}
\caption{Color--magnitude diagram for the photometric and spectroscopic catalogs of the A1240 field. 
Black scatters show Subaru photometric catalog. 
Blue squares represent galaxies spectroscopically confirmed as members of A1240 and A1237 (the bottom panel of Figure~\ref{fig:speczhist}), while orange crosses indicate foreground plus background galaxies. 
The locus of red-sequence galaxies is well-defined with confirmed member galaxies. 
We applied its color--magnitude relation (dashed line) to identify red-sequence member candidates (red points) from the photometric catalog. 
The source galaxies selected for the WL analysis are shown as a green Hess diagram with square-root scaling (see \textsection\ref{subsubsec:srcgal} for details).
}\label{fig:cmd}
\end{center}
\end{figure}

Although we tripled the number of the A1240 spectroscopic members, the spectroscopic catalog is still far from complete. Thus,
we used the photometric catalog from Subaru/Suprime-Cam to augment our spec-$z$ member catalog and create the final cluster galaxy catalog.
Since the $g-r$ color brackets the redshifted 4000~\AA\ break---the characteristic spectral feature caused by the dominant old, cool stellar populations---at $z=0.195$, 
its red-sequence is readily distinguished in the color--magnitude diagram (CMD). 
Figure~\ref{fig:cmd} shows the $g-r$ vs. $r$ CMD of the A1240 field.
The locus of the red-sequence is well-defined photometrically and also nicely traced by the spectroscopically confirmed members. 
The color--magnitude relation of the red-sequence is determined from the best-fit linear regression of the spectroscopic members within the $g-r$ color range of $1.079\pm0.187$, which is computed with the bi-weight estimator. 
The resulting best-fit result is $g-r = -0.032r+1.723$.
This relation is used for identifying photometric red-sequence galaxies. 
We restricted the length of the red sequence to objects six magnitudes fainter than the brightest cluster galaxy (BCG)\footnote{Following our previous \mcc\ work \citepalias{Golovich+2019b}, we determined the width of the red-sequence box to be six magnitudes, sufficient to cover the magnitude range of spectroscopically confirmed member galaxies.} and width to $\pm1\sigma$, where $\sigma$ is a robust standard deviation of the $g-r$ colors of the confirmed member galaxies.

\begin{figure}
\begin{center}
\includegraphics[width=1\linewidth]{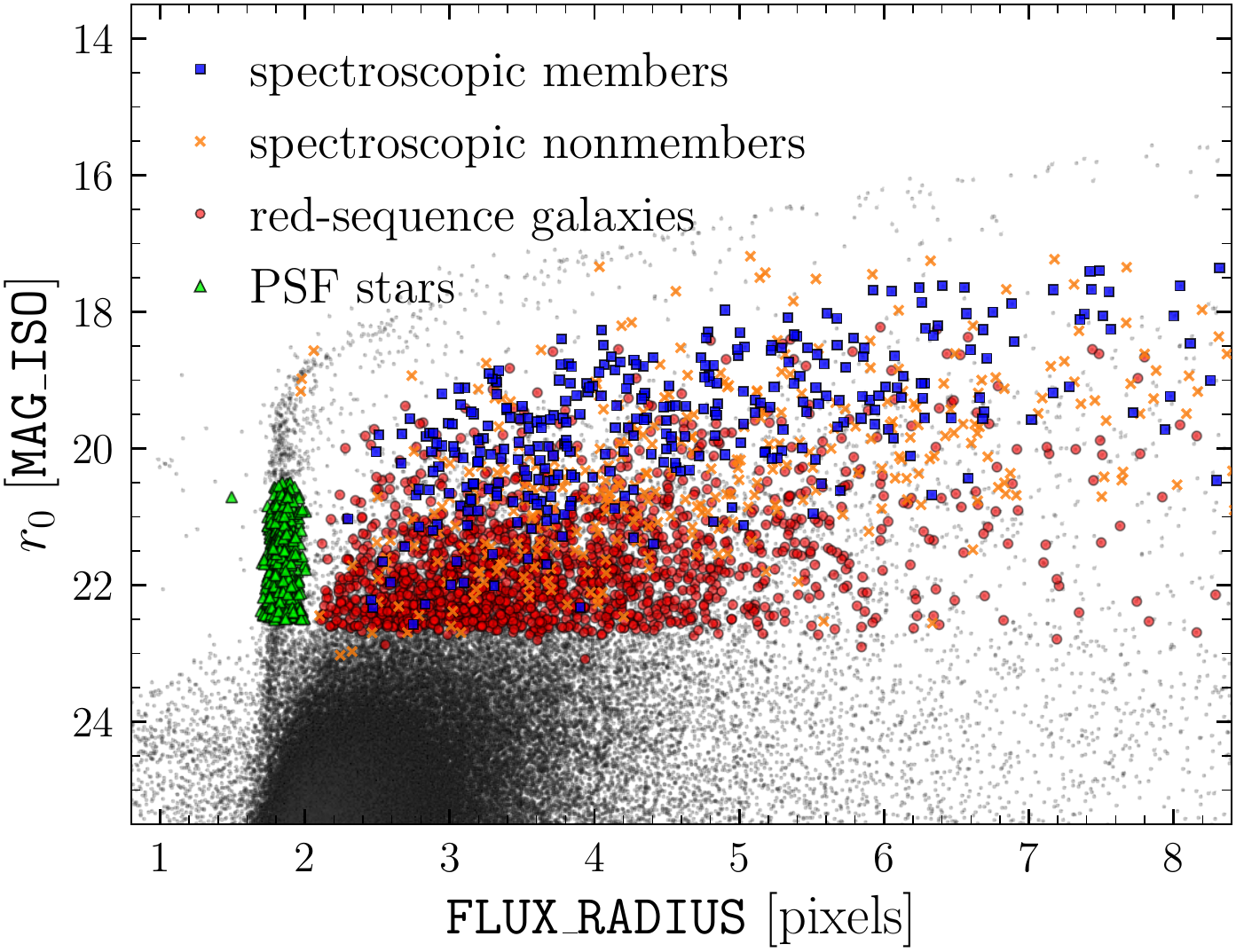}
\caption{Size--magnitude diagram for the photometric and spectroscopic catalogs of the A1240 field. 
The stars follow a tight locus around a half-light radius ({\tt FLUX\_RADIUS} from SExtractor) of 1.8 pixel. 
The point-like objects, even saturated brighter stars, are clearly distinguished from the spectroscopically confirmed members (blue squares). 
The stellar size--magnitude relations are used for separating stars and photometric red-sequence galaxies, along with the color--magnitude relation in Figure~\ref{fig:cmd}. 
The green triangles indicate stars selected for the PSF modeling in \textsection\ref{subsubsec:psfmodel}.
}\label{fig:sizemag}
\end{center}
\end{figure}

\begin{figure*}
\includegraphics[width=0.469\linewidth]{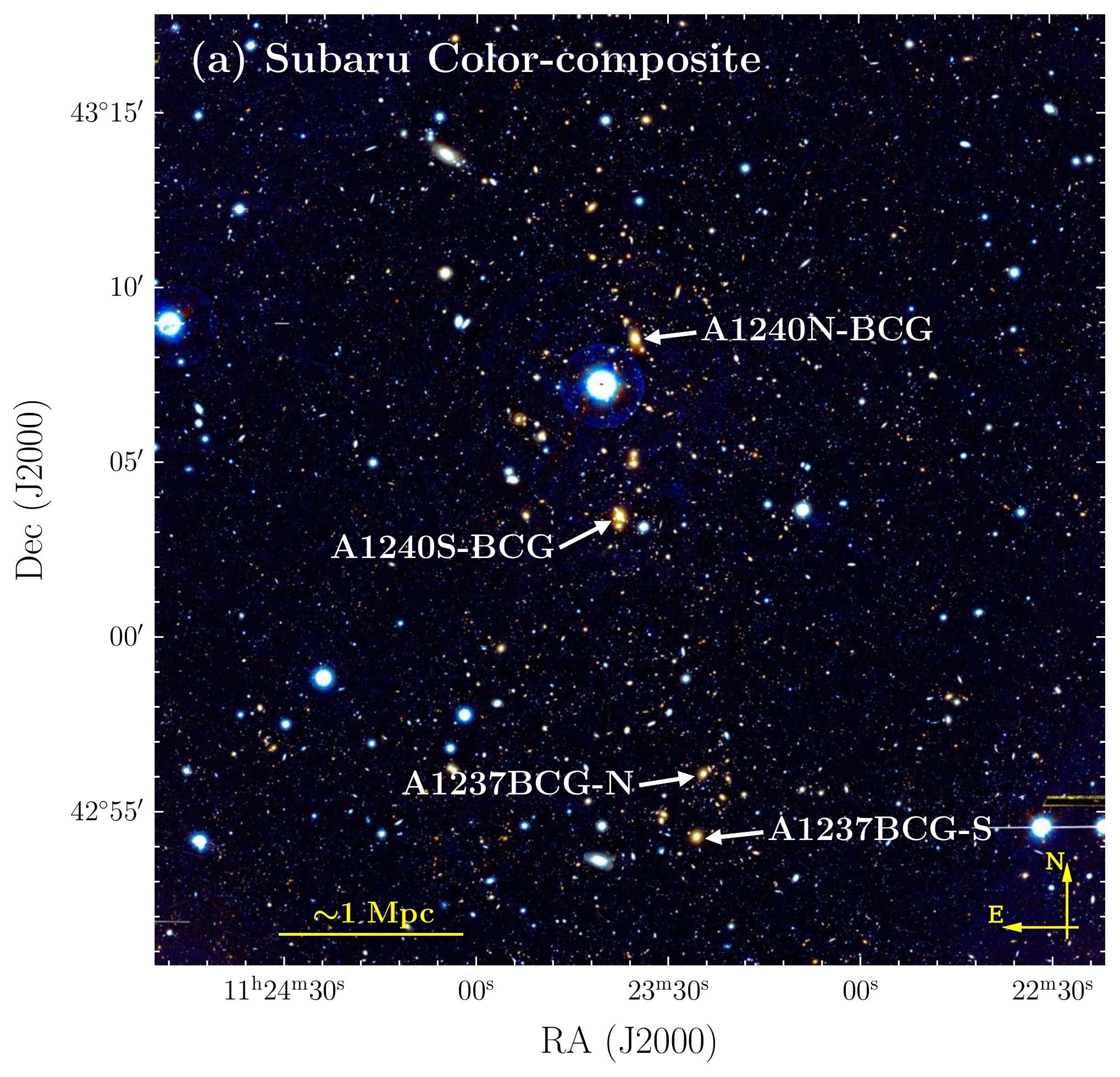}
\includegraphics[width=0.521\linewidth]{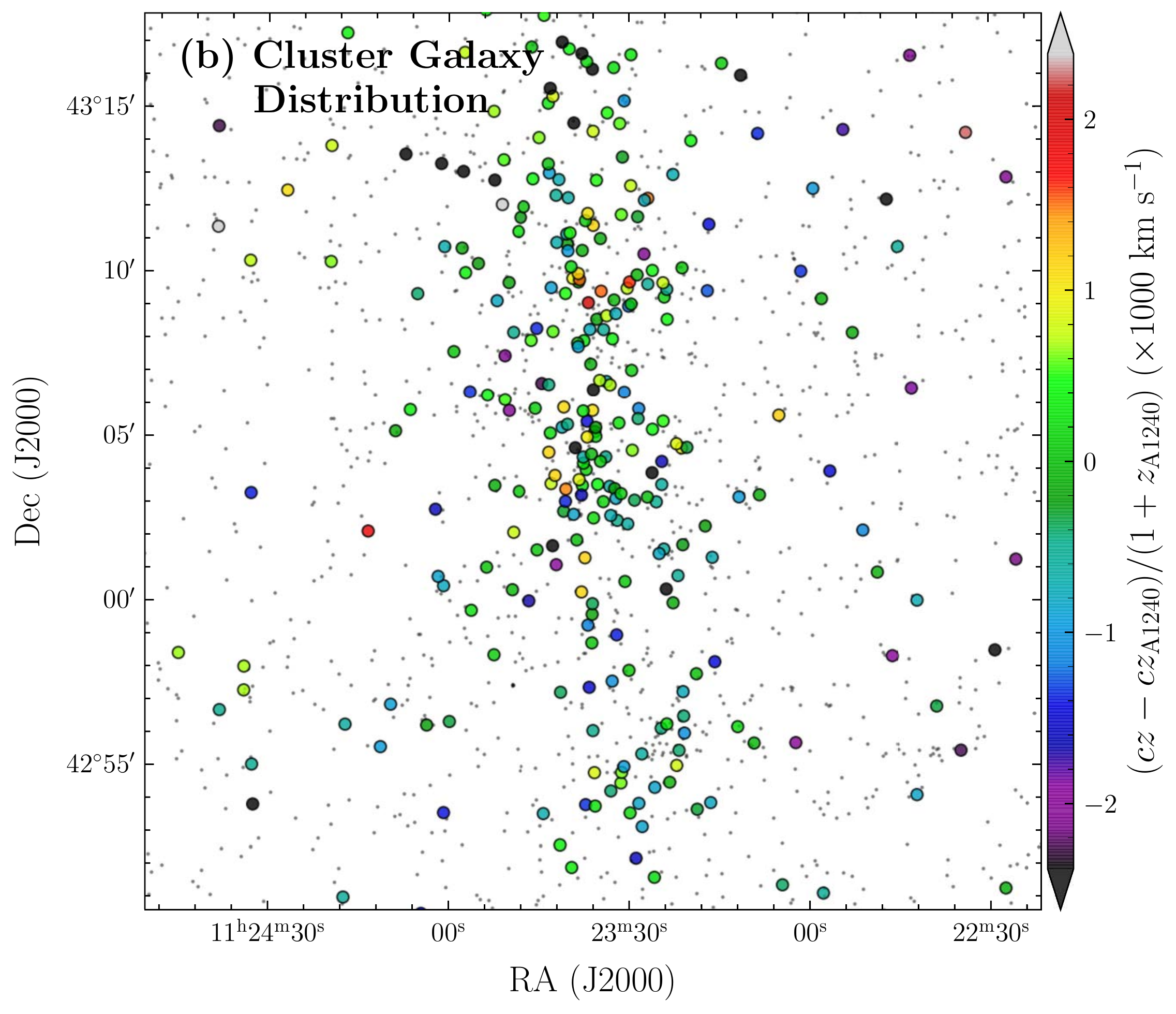}
\includegraphics[width=0.469\linewidth]{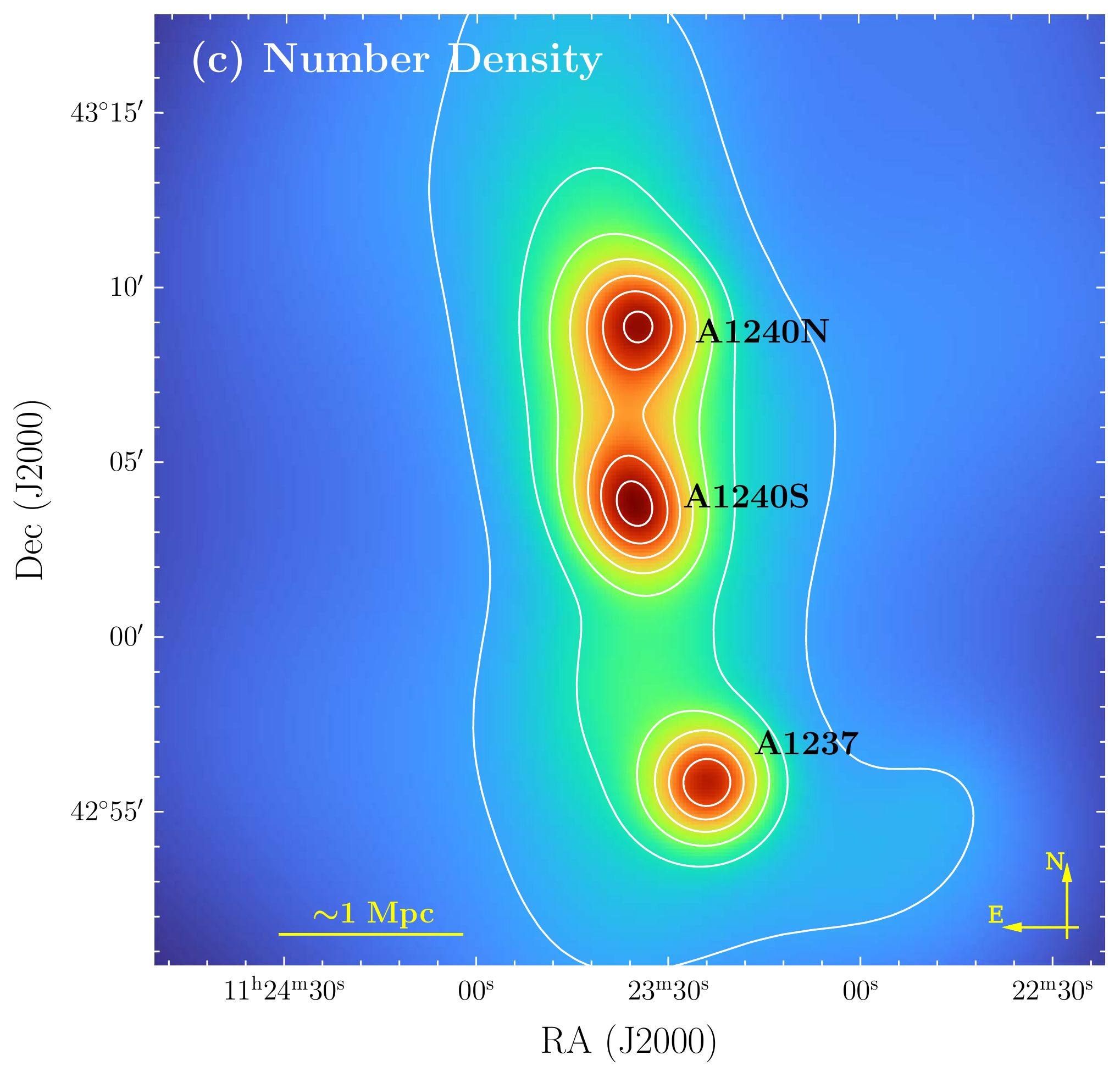}
\includegraphics[width=0.469\linewidth]{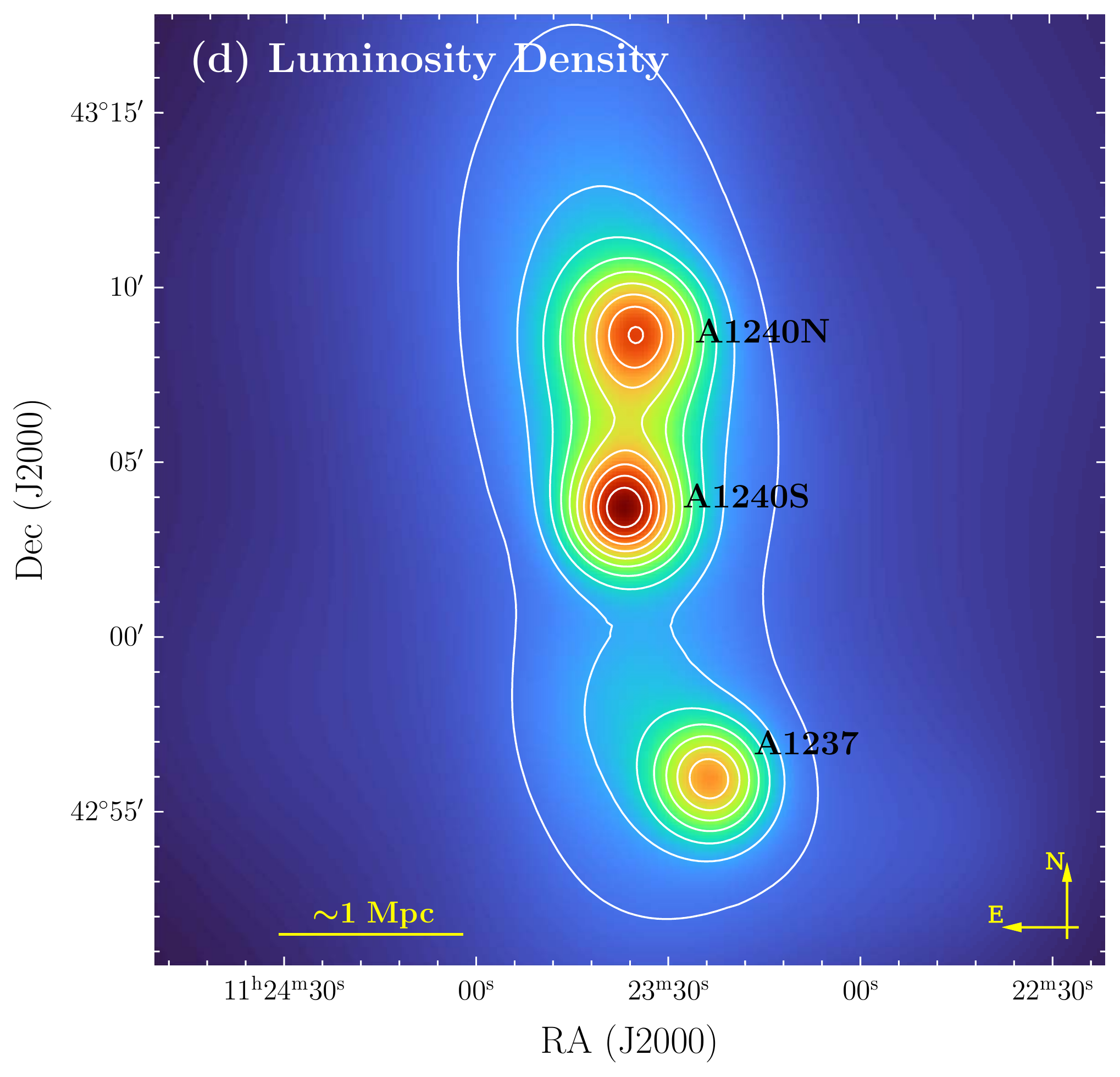}
\caption{A1240 optical image and cluster member distribution.
{\it Top left}: (a) color-composite image of the central $27\arcmin\times27\arcmin$ (5.24~Mpc $\times$ 5.24~Mpc) region.
The Subaru/Suprime-Cam $r$, $g+r$, and $g$-band data are used for the RGB channels of the image.
{\it Top right}: (b) distribution of the A1240$+$A1237 member (candidate) galaxies. 
Filled circles show the locations of the spectroscopically confirmed cluster members, color-coded with the relative velocity difference with respect to the cluster mean velocity
of A1240.
Gray points are the photometrically selected (red-sequence) members.
{\it Bottom left}: (c) adaptively-smoothed galaxy number density map. 
We used the CIAO {\tt csmooth} tool with a minimum (maximum) significance of 3$\sigma$ (5$\sigma$).
The resulting minimal smoothing scale is $60\arcsec$ ($\approx0.194$~Mpc at the cluster redshift). 
{\it Bottom right}: (d) adaptively-smoothed luminosity-weighted galaxy density map. To smooth the luminosity map, we reused the kernel scale map that {\tt csmooth} generated for the number density map.
In the bottom panels, the contours begin at $1\sigma$ above mean density and are linearly spaced in units of $\sigma$, where $\sigma$ is the standard deviation of density distribution. 
The galaxy distribution shows that A1240 consists of two subclusters (A1240N and A1240S) separated by \mytilde0.96 Mpc. A1237, \mytilde1.56~Mpc to the south of A1240S, is detected
as a compact distribution of galaxies in both luminosity and number densities with a mean redshift similar to that of A1240.
}\label{fig:galmap}
\end{figure*}

In addition to the above selection criteria in the color--magnitude space, we employed the size--magnitude relation in eliminating stars from the photometric red-sequence catalog. 
Figure~\ref{fig:sizemag} shows the half-light radius ({\tt FLUX\_RADIUS}) vs. $r$ ({\tt MAG\_ISO}) relation for the same objects in Figure~\ref{fig:cmd}.
The point-like sources and bright saturated stars are either more compact or brighter than the confirmed member galaxies. 
We merged the resulting red-sequence catalog and the spectroscopic cluster galaxy catalog to investigate the projected surface density distributions of the cluster galaxies described below.

Figure~\ref{fig:galmap} shows the cluster galaxy distributions in the central $27\arcmin\times27\arcmin$ (5.24~Mpc $\times$ 5.24~Mpc) region of the A1240 field. 
In the panel (a), we display the color-composite image, created from Subaru/Suprime-Cam $r$, $g+r$, and $g$-band data. We mark the positions of the BCGs in A1240 and A1237. 
We refer to two BCGs of A1240 as A1240N-BCG and A1240S-BCG, which are located at the approximate centers of the northern (A1240N) and southern (A1240S) subclusters in A1240.
Our current data do not show distinct substructures in A1237. The two BCGs in A1237 are labeled as A1237BCG-N and A1237BCG-S according to their location in the Subaru image.
In the panel (b), we display both spectroscopic (circle) and photometric (gray dot) members. We color-coded spectroscopic member galaxies using the relative velocity difference with respect to the rest-frame central velocity of A1240. 

To detect density peaks and substructures, we adaptively smoothed the two-dimensional (2D) density distribution of cluster galaxies. 
The surface number density map was produced by applying the CIAO\footnote{\url{https://cxc.harvard.edu/ciao4.13/}} \citep{ciao} {\tt csmooth} tool with a minimal smoothing scale of 1\arcmin\ ($\approx0.194$~Mpc at $z=0.195$) and a significance between 3$\sigma$ and 5$\sigma$. 
The resulting map is displayed in the panel (c). 
The panel (d) shows the adaptively-smoothed density map weighted by luminosity (SExtractor isophotal flux {\tt FLUX\_ISO} parameter). 
We here reused the kernel size map output by {\tt csmooth} during the creation of the number density map.

Both the number and luminosity-weighted density maps clearly identify three overdense regions aligned in the N--S direction.
A1240 consists of two subclusters (A1240N and A1240S) separated by \mytilde1~Mpc.
This bimodality is consistent with the findings of \citetalias{Barrena+2009} and \citetalias{Golovich+2019b}.
About \mytilde1.6~Mpc to the south of A1240S is A1237, which is also shown in
\citetalias{Barrena+2009} and \citetalias{Golovich+2019b}.

\begin{figure}
\centering
\includegraphics[width=1.0\linewidth]{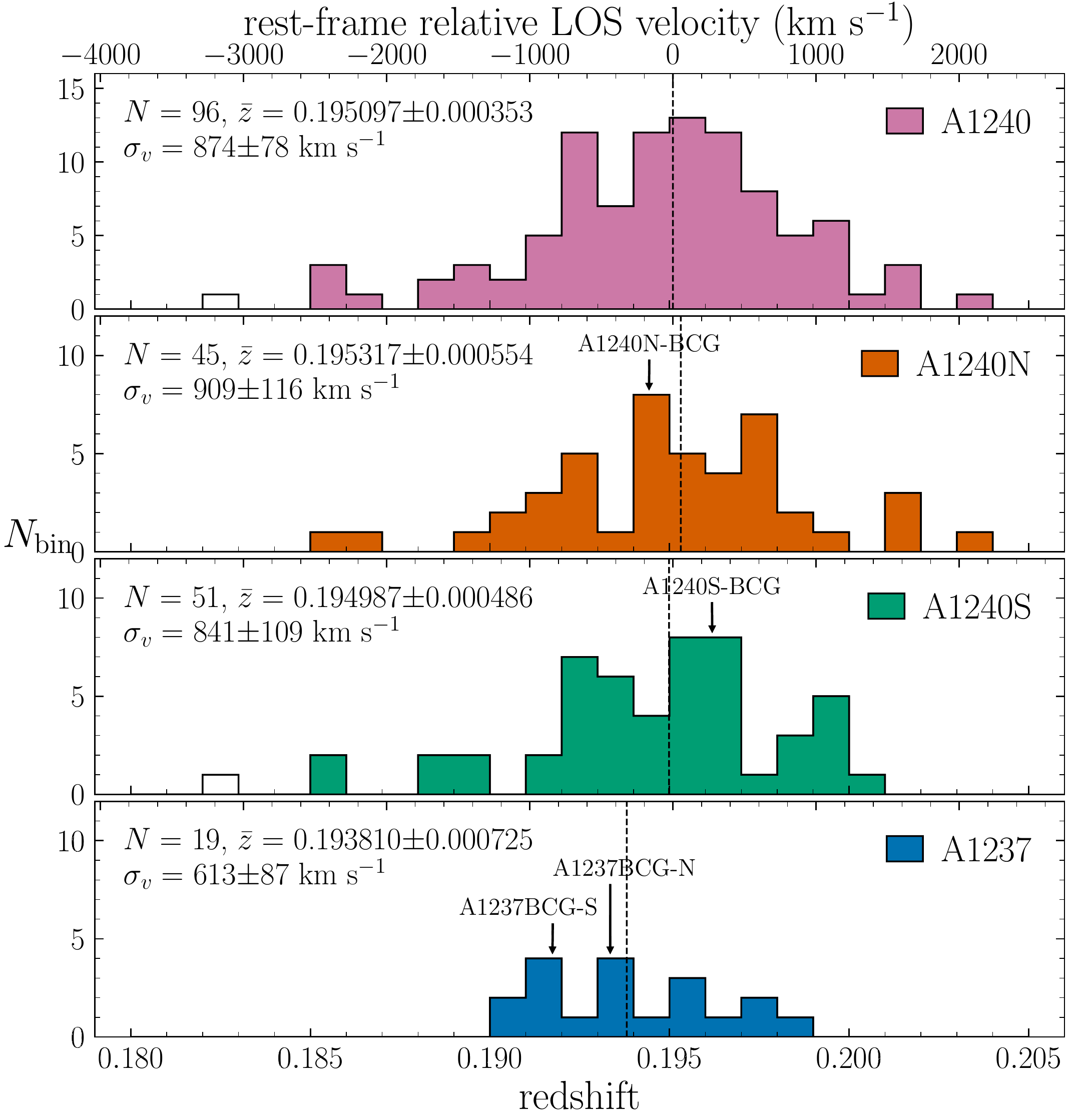}
\caption{Redshift distributions of luminosity peaks. 
The top panel shows the combination of A1240N and A1240S, while the lower panels are redshift distributions of galaxies belonging to individual luminosity peaks within a radius of $\mytilde2\farcm5$ (\mytilde0.48~Mpc). The bi-weight estimators are applied to measure mean redshifts (dashed vertical lines) and rest-frame velocity dispersions with measurement uncertainties extracted from the bootstrap resampling. The empty histograms represent the galaxies excluded from the bi-weight statistics presented here. The arrows denote the spectroscopic redshifts of BCGs. The velocity scale at the top axis is centered on the A1240 rest-frame central velocity of $z=0.195097$ from the top panel. 
}\label{fig:speczdist3sig}
\end{figure}

Figure~\ref{fig:speczdist3sig} shows the redshift distribution of the cluster galaxies belonging to each substructure.
Because of the small differences in the LOS velocities and the overlapping virial radii, it is difficult to assign a unique membership to each galaxy.
Thus, we relied on the projected separation from the substructure center to determine the substructure membership, labeling an object
as a member if it is within $\mytilde2\farcm5$ (half of the distance between A1240N and A1240S, i.e., \mytilde0.48~Mpc).
To define the substructure center, we used the luminosity-weighted density map. We note that similar results are obtained when the number density map is used instead.
The resulting numbers of objects are 45, 51, and 19 for A1240N, A1240S, and A1237.

The LOS velocity difference between A1240N and A1240S is estimated to be smaller ($83\pm185$~\kms) than our previous estimate ($394\pm117$~\kms) from \citetalias{Golovich+2019b}.
With respect to A1240S, A1237 is offset by $295\pm219$~\kms, which is in agreement with the estimates of \citetalias{Barrena+2009} ($59\pm203$~\kms) and \citetalias{Golovich+2019b} ($88\pm159$~\kms) within the errors. 
Therefore, our analysis suggests that the LOS velocity
difference among the three substructures is consistent with zero.

Velocity dispersions are computed with the bi-weight estimator. We find $909\pm116$~\kms, $841\pm109$~\kms, and $613\pm87$~\kms\ for A1240N, A1240S, and A1237, respectively. Based on GMM analysis (including more member galaxies), \citetalias{Golovich+2019b} quoted $706\pm52$~\kms\ and $727\pm68$~\kms\ for A1240N and A1240S, respectively, which are broadly consistent with the current estimates. Our velocity dispersion for A1237 is also consistent with the \citetalias{Barrena+2009} estimate ($738_{-54}^{+82}$~\kms).
Detailed discussions on the comparison with the previous studies are presented in \textsection\ref{sec:discussion}.
For each subcluster, the 2D density peak coordinates, redshift and velocity dispersion measurements, and their bootstrapped uncertainties are summarized in Table~\ref{tab:subclusters}.
Note that sometimes an incorrect value of $z=0.159$ is quoted for the redshift of A1240 in the literature \citep[e.g.,][]{Kempner+Sarazin2001,Bonafede+2009,Feretti+2012review,deGasperin+2014_PSZ1G096.89+24.17,vanWeeren2019review,Wittor+2021}. In fact, the value $z=0.159$ was estimated from the apparent magnitude of the tenth brightest cluster galaxy in A1240 \citep{David+1999}.

\subsubsection{Filamentary Structure around A1240} \label{subsubsec:filament}

Numerical simulations have shown that galaxy clusters grow by accreting galaxies, groups, or clusters along their host filaments. Therefore, one of the natural expectations is that cluster merger axes might preferentially align with the orientations of the filaments. 
Our galaxy analysis with Subaru and MMT data show that the three substructures in the A1240 field 
form a \mytilde4~Mpc filamentary structure 
in the N--S orientation.
Here, we investigate the surrounding environments of A1240 out to radii of \mytilde100~Mpc from the cluster center and report a discovery of a much larger-scale (\mytilde80~Mpc) filamentary structure whose orientation is remarkably consistent with the hypothesized merger axis of A1240.

\begin{figure*}
\centering
\includegraphics[width=1.0\linewidth]{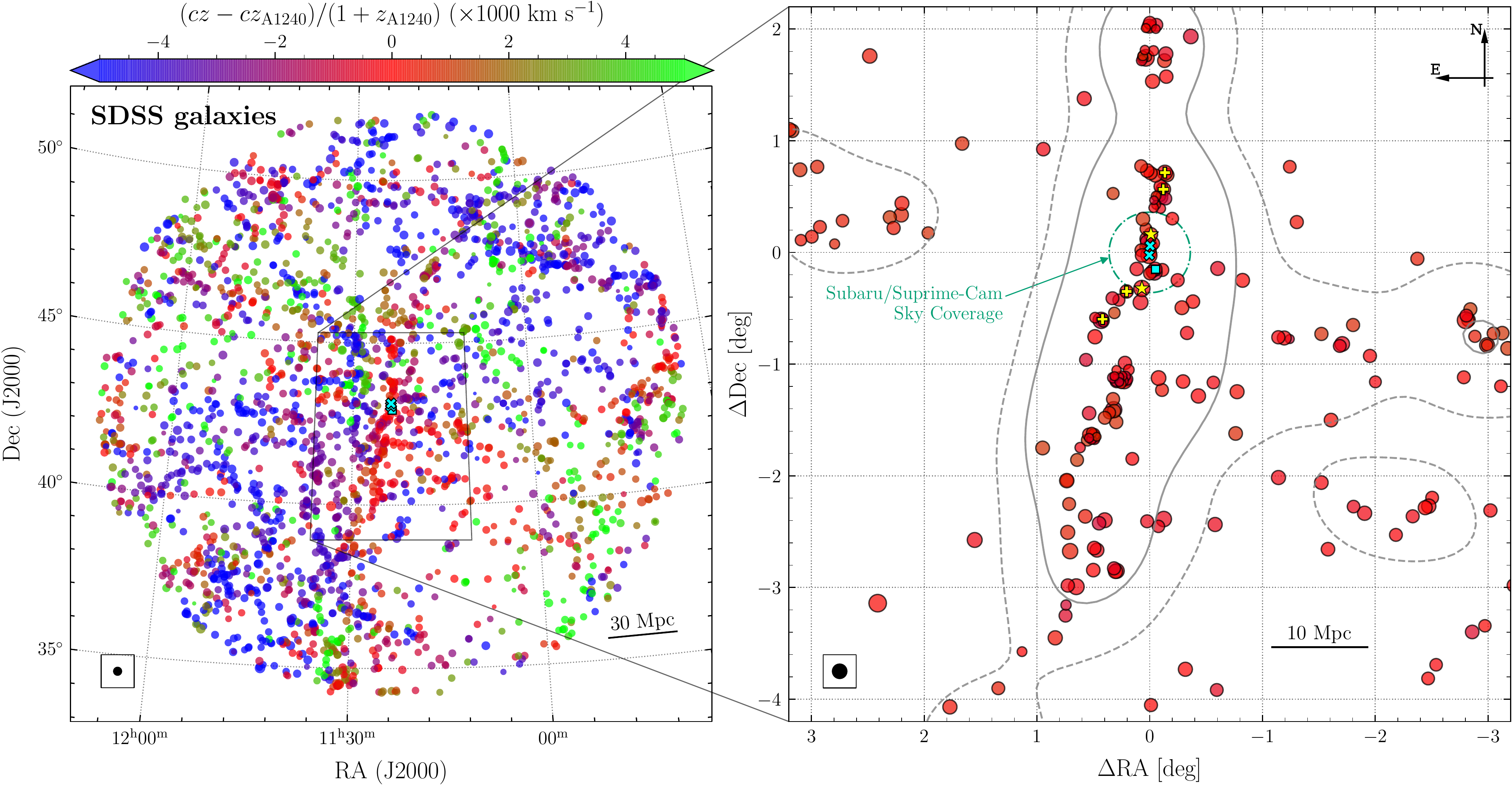} 
\caption{Spatial distribution of the SDSS DR16 galaxies around A1240. 
The left panel displays the $\mytilde3000$ objects at the clustercentric radius $R < 9\degr$ (\mytilde100~Mpc) whose rest-frame LOS velocities are within $\pm6000$~\kms\ of the cluster.
The symbols are color-coded according to their velocities
relative to the redshift of A1240.
We let the symbol size scale with the dereddened SDSS $r$-band magnitude. The size of the black filled circle in the lower-left corner box represents the A1240 BCG magnitude.
A zoomed-in view of the central box is shown in the right panel, where we only show the galaxies within the velocity range $\pm1000$~\kms. 
The green dash-dotted circle marks the sky coverage of the current Subaru/Suprime-Cam data. 
The yellow symbols indicate the positions of the (candidate) clusters of galaxies, inside ($R\lesssim20\arcmin$, {\it star}) and outside ($20\arcmin \lesssim R \lesssim 54\arcmin$, {\it plus}) of the Subaru coverage, compiled from the literature.
We created a galaxy number density map by smoothing this with a FWHM = 5~Mpc Gaussian kernel and illustrate the $3\sigma$ and $1\sigma$ levels above the mean number density
with gray solid and dashed contours, respectively.
In both panels, cyan symbols mark the locations of the luminosity density peaks of A1240 ({\it cross}) and A1237 ({\it square}). 
Remarkably, A1240 is embedded within the much larger \mytilde80~Mpc filamentary structure elongated in the N--S orientation, which is in good agreement with the hypothesized merger axis of A1240. Also, note that the six known confirmed/candidate clusters together with the three substructures within the Subaru field form a $\mytilde20$~Mpc-long filament in the same N--S orientation. 
The fact that both ends of the $\mytilde80$~Mpc-long filament do not differ greatly in their LOS velocities suggests that the filament may be nearly aligned with the plane of the sky.
}\label{fig:filament}
\end{figure*}

We retrieved a galaxy catalog for the radius of $R < 9\degr$ ($\mytilde100$~Mpc at $z=0.195$) region centered on A1240 from the SDSS DR16 archive. In addition, we compiled a list of galaxy clusters at $R < 0.9\degr$ (\mytilde10~Mpc) 
whose redshifts are similar to that of A1240 from the literature \citep{Abell1958, Abell1989, MaxBCG2007, GMBCG2010, WHL12, WH15}\footnote{
For example, a group of galaxies \mytilde1.9~Mpc north of A1240 was identified as a cluster by \citet[][GMBCG J170.88991+43.24646, hereafter GMBCG170.89]{GMBCG2010} and \citet[][WHL J112333.6+431447]{WHL12}. Another example is WHL J112352.7+424542 \citep{WHL12} detected \mytilde2.3~Mpc to the southeast of A1237. These two clusters are marked with yellow stars in the right panel of Figure~\ref{fig:filament}.
}. 
The result is displayed in Figure~\ref{fig:filament}. Remarkably, we find that A1240 is embedded within a much larger filament stretched out to $\mytilde80$~Mpc whose orientation is in good agreement with the hypothesized merger axis.
Also, the six known (candidate) clusters together with the three substructures within the Subaru field form a $\mytilde20$~Mpc-long filament in the same N--S orientation. 
The relative rest-frame LOS velocities $\left| \Delta v_\mathrm{rf}\right|$ of these clusters are less than \mytilde300~\kms\ except for the southernmost A1253 with $\Delta v_\mathrm{rf}\sim1100$~\kms. Moreover, the northern and southern ends of the $\mytilde80$~Mpc-long filament do not differ greatly in their LOS velocities. These facts suggest that the filament around A1240 may be nearly aligned with the plane of the sky.
It will be interesting to investigate the cosmic filament detected in the current study with WL based on future wider and deeper imaging data.

\begin{deluxetable*}{cccCLCC}
\tablecaption{Measured Properties of Subclusters in the A1240 Field\label{tab:subclusters}}
\tablewidth{0pt}
\tablehead{
\multicolumn1c{} & \multicolumn1c{Number Density Peak}\vspace{-0.25cm} & \multicolumn1c{Luminosity Density Peak} & \multicolumn1c{} & \multicolumn1c{$\sigma_{v}$} & \multicolumn1c{$\sigma_{v,\rm SIS}$} & \multicolumn1c{$M_{200}$} \\
\multicolumn1c{Subcluster}\vspace{-0.25cm} & & & \multicolumn1c{$z$} & \multicolumn1c{} &  & \\
\multicolumn1c{} & \multicolumn1c{R.A., Decl. (deg)} & \multicolumn1c{R.A., Decl. (deg)} & \multicolumn1c{} & \multicolumn1c{(\kms)} & \multicolumn1c{(\kms)} & \multicolumn1c{($10^{14}$\Msun)} 
}
\decimalcolnumbers
\startdata
A1240N & 170.894181, 43.148796 & 170.896484, 43.143749 & 0.195317\pm0.000554 & 909\pm116 & 540_{-55}^{+50} & 2.61_{-0.60}^{+0.51} \\
A1240S & 170.896442, 43.062998 & 170.903349, 43.061313 & 0.194987\pm0.000486 & 841\pm109 & 464_{-63}^{+55} & 1.09_{-0.43}^{+0.34} \\
A1237 & 170.850419, 42.930099  & 170.848120, 42.933463 & 0.193810\pm0.000725 & 613\pm87 & 447_{-68}^{+59} & 1.78_{-0.55}^{+0.44} \\
\enddata
\tablecomments{The columns list: (1) subcluster component; (2) and (3) coordinate of the number and luminosity-weighted density peaks (see bottom panels of Figure~\ref{fig:galmap}); (4) and (5) mean redshift and velocity dispersion of luminosity peaks (see Figure~\ref{fig:speczdist3sig}); (6) velocity dispersion determined from the best-fit singular isothermal sphere (SIS) profile (see \textsection\ref{subsubsec:comp-kin}); and (7) WL mass estimate based on the best-fit NFW profile (see \textsection\ref{subsubsec:massfit}).}
\end{deluxetable*}

\subsection{Weak Lensing} \label{subsec:wlanal}

\subsubsection{Basic Theory} \label{subsubsec:wlbasic}

In this section, we provide a brief overview of WL theory and its formalism. For details, we refer the reader to review papers \citep[e.g.,][]{Narayan+1996review,WLreviewBS01,Kneib+2011review}.

Galaxy clusters act as a gravitational lens for background galaxies. 
In the WL regime where a source galaxy is much smaller than the characteristic scale of the lensing signal variation, the coordinate transformation by gravitational lensing can be linearized. 
The resulting Jacobian matrix $\mathbf{A}$ transforming the source plane position $\mathbf{x}$ to the image plane position $\mathbf{x^{\prime}}$ via $\mathbf{x^{\prime}}=\mathbf{A} \mathbf{x}$ is then described by:
\begin{equation}
    \mathbf{A}= (1-\kappa)
    \begin{pmatrix}
    1-g_1  & -g_2 \\
    -g_2 & 1 + g_1
    \end{pmatrix},  \label{eqn:A}
\end{equation}
where $\kappa$ is the convergence and
$g_{1(2)}$ is the first (second) component of the reduced shear $g$: 
\begin{equation}
	g=(g_1^2+g_2^2)^{1/2}\equiv\gamma/(1-\kappa). \label{eqn:g}
\end{equation}
In Equation~(\ref{eqn:g}), 
$\gamma$ is the shear, which approaches $g$ when $\kappa \ll 1$.
The convergence $\kappa$ is the dimensionless surface mass density defined as the ratio of the projected surface mass density $\Sigma$ and the critical surface mass density $\Sigma_{c}$: 
\begin{equation}
\kappa=\frac{\Sigma}{\Sigma_c}. \label{eqn:kappa}
\end{equation}
\noindent
The critical surface density $\Sigma_{c}$ is defined by
\begin{equation}
\Sigma_c=\frac{c^2}{4\pi G}\frac{D_{s}}{D_l D_{ls}}, \label{eqn:sigma_c}
\end{equation}
where $c$ is the speed of light, $G$ is the gravitational constant, 
and $D_l$, $D_{ls}$, and $D_s$ are the angular diameter distances from the observer to the lens, from the lens to the source, and from the observer to the source, respectively.

The matrix $\mathbf{A}$ in Equation~(\ref{eqn:A}) transforms a circular source into an ellipse. 
To define the ellipticity, we model an object with an elliptical Gaussian whose semi-major and -minor axes are $a$ and $b$, respectively. 
The reduced shear $g$ then becomes
\begin{equation}
g=\frac{a-b}{a+b}. \label{eqn:e_definition}
\end{equation}
As the ellipse also has an orientation, one can conveniently represent both its magnitude and position angle using the complex notation:
\begin{equation}
\mathbf{g}=g_1 + \mathbf{i} g_2 \equiv g e^{2i\phi}, \label{eqn:complexg}
\end{equation}
with magnitude $g$ in Equation~(\ref{eqn:g}) and orientation angle $\phi=0.5 \tan^{-1}(g_2/g_1)$ of the semi-major axis.

In the same manner, the unlensed (intrinsic) ellipticity of the source galaxy can be formulated as
\begin{equation}
\boldsymbol{\epsilon}=\epsilon_1 + \mathbf{i} \epsilon_2. \label{eqn:complexe}
\end{equation}
Then, the transformation between the source galaxy intrinsic ellipticity $\boldsymbol{\epsilon}$ and the lensed (observed) ellipticity $\mathbf{e}$ is given by 
\begin{equation}
\mathbf{e} = \frac{ \boldsymbol{\epsilon} + \mathbf{g}} { 1 + \mathbf{g}^* \boldsymbol{\epsilon}}, \label{eqn:e_transform}
\end{equation}
where the asterisk superscript denotes the complex conjugate.
In a typical WL regime where $\kappa\ll1$, $\gamma\ll1$, and thus $g\ll1$, 
in general each galaxy's lensed ellipticity $\mathbf{e}$ is slightly offset from its intrinsic ellipticity $\boldsymbol{\epsilon}$. 
Assuming isotropic distribution of $\boldsymbol{\epsilon}$, one can obtain $\mathbf{g}$ via
\begin{equation}
\mathbf{g} = \left < \mathbf{e} \right > \label{eqn:average}
\end{equation}
when no bias is present. 

\subsubsection{PSF Modeling} \label{subsubsec:psfmodel}

In the ground-based observation, the atmospheric turbulence and instrumental imperfection cause variations in size and ellipticity of the PSF. The PSF, on average, dilutes the observed shear of source galaxies whereas locally it induces anisotropic bias on galaxy ellipticity measurements.
In order to recover the intrinsic lensing signal, it is crucial to correct for the PSF-induced dilution and anisotropy by carefully modeling the PSF. 
We modeled the PSF based on the principal component analysis (PCA) technique presented in \citet{J07PCA}. Here we provide a brief overview and refer the reader to \citet{J07PCA}, \citet{JT11PCA}, and \citet{Jee13DLS} for details.

The first step in modeling the PSF variation is to identify ``good'' stars from each CCD frame. We selected stars on individual resampled frames ({\tt RESAMP}) produced by SWarp as a part of the two-step co-adding procedure, using SExtractor and the size--magnitude relations. The average number of stars per frame suitable for PSF modeling is \mytilde30. We note that the density of stars in this field is lower than our previous \mcc\ WL studies (typically containing $\ga 100$ per CCD) because of the high Galactic latitude of A1240 ($b=66\degr$). Next, we determined the mean PSF per {\tt RESAMP} and the deviations from the mean for individual stars by median-stacking the postage stamp cutouts (21~pixel $\times$ 21~pixel) and subtracting the median from the cutouts. We performed PCA on the deviations to obtain the covariance matrix and compute eigenvalues and the corresponding eigenvectors (i.e., principal components) of the matrix. We then fitted third-order polynomials to the amplitudes along the eigenvectors to obtain a model PSF at any arbitrary position on each {\tt RESAMP}. Finally, we stacked the model PSFs from the individual {\tt RESAMP} images to create a coadd PSF $P(x,y)$ at a galaxy position on the co-add image.

\begin{figure*}
\centering
\includegraphics[width=1\linewidth]{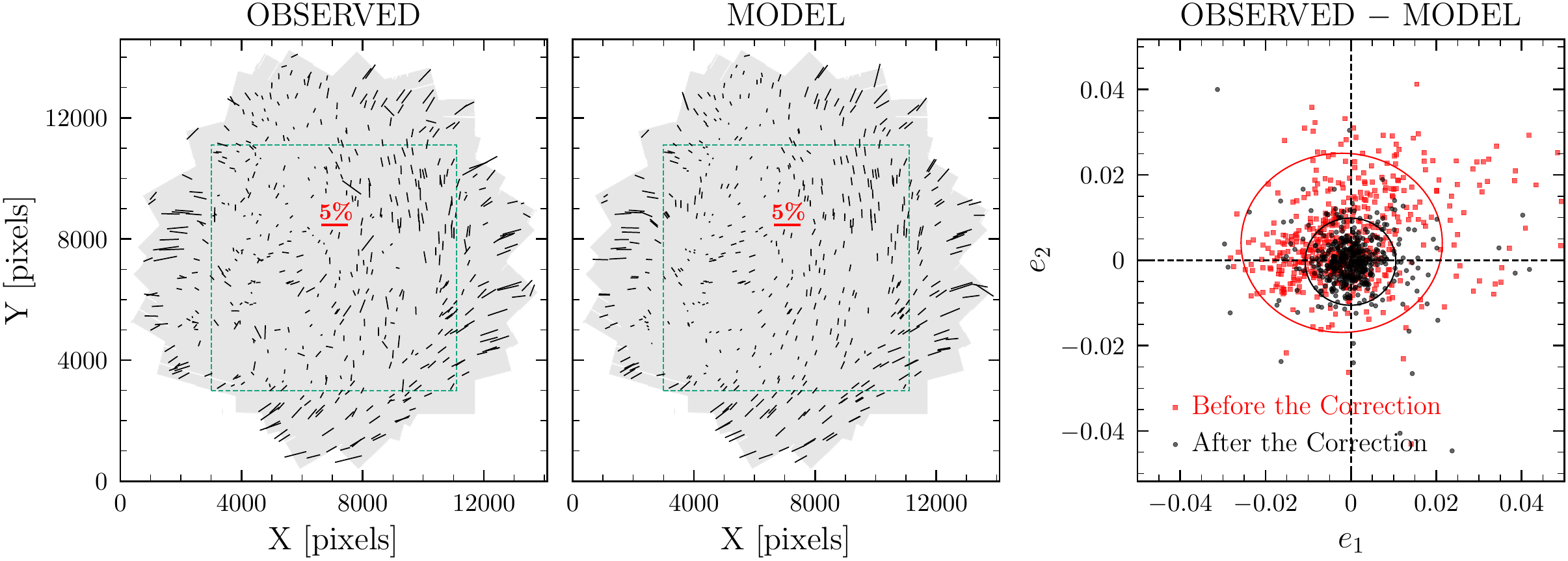}
\caption{Observed PSF vs. model PSF. 
The left and middle panels show the measured ellipticity distributions of observed stars and model PSFs in the Subaru $r$-band co-add image. 
The length and direction of ``whiskers'' are proportional to the magnitude and orientation of the ellipticities, respectively. 
The corresponding model PSFs are in good agreement with the observed PSF ellipticities except for several outliers. 
The red whiskers show the size of 5\% ellipticity.
The green dashed boxes mark the central $27\arcmin\times27\arcmin$ region, where we performed 2D galaxy distribution mapping and WL mass reconstruction. 
The right panel displays the ellipticity components of the observed stars (red squares) and the residuals after applying PSF correction (black points). The ellipses, with the same color scheme, characterize the distributions of ellipticity components. 
The center and radii of each ellipse are defined by the bi-weight locations and $2\sigma$ in both ellipticity component axes, where $\sigma$ is the bi-weight scale. 
The correction reduces the size of the scatters and also makes the centroid of the data points closer to the origin
[$(0.0022\pm0.0005,0.0041\pm0.0005) \rightarrow (-0.0001\pm0.0002,-0.0003\pm0.0002)$].
}\label{fig:psfmodel}
\end{figure*}

The left panel of Figure~\ref{fig:psfmodel} displays the ellipticities directly measured from \mytilde500 stars (shown as the green triangles in the size--magnitude diagram of Figure~\ref{fig:sizemag}) detected in the co-add frame. The stellar PSF patterns vary spatially across the entire field.
For a visual comparison, we display the ellipticities of the PSFs predicted at the same positions of the stars (middle panel). 
The ellipticity variation pattern of the PSFs obtained through our PCA-based model shows good agreement with that of the observed stars on the co-add frame. Some apparent outliers are attributed to cosmic-ray affected stars, galaxies misclassified as stars, stars severely blended with nearby bright sources, or saturated trails. 
To examine how well the model PSFs describe the observed ellipticities quantitatively, we compare the ellipticity components before (red) and after (black) the correction in the right panel of Figure~\ref{fig:psfmodel}. The correction not only reduces the size of the scatters (compare the red and black ellipses), but also makes the centroid of the data points closer to the origin
[$(0.0022\pm0.0005,0.0041\pm0.0005) \rightarrow (-0.0001\pm0.0002,-0.0003\pm0.0002)$].

\subsubsection{Shape Measurement} \label{subsubsec:shape}

Reduced shears $g_{1(2)}$ are derived by averaging over the ellipticities of source galaxies (Eqn.~\ref{eqn:average}).
In real observations, however, the averaged ellipticities are only biased measures of reduced shears. Sources of the systematics include model/underfitting, noise, PSF model, and pixellation biases \citep[e.g.,][]{Bernstein2010,MV2012,Refregier+2012,Great3+2014}.
In the current study, we choose to address the issues with the ``forward-modeling'' and WL image simulation approaches.

For each object detected with SExtractor, we first created a square postage-stamp cutout centered on the object $(x_\mathrm{c}, y_\mathrm{c})$ with $(8w+20)$ pixels on a side, where $w$ is the SExtractor's semi-major axis {\tt A\_IMAGE} measured from the $r$-band mosaic image. 
The pixels belonging to the neighboring objects were masked out using the SExtractor segmentation map. 
We then convolved a 2D elliptical Gaussian function $G(x,y)$ by a model PSF profile $P(x,y)$ via Fast Fourier Transform:
\begin{equation}
	M(x,y) = G(x,y) \otimes P(x,y), \label{eqn:model}
\end{equation}
where
\begin{equation}
	\label{eqn:egauss}
	\begin{aligned}
		G(x,y) = I_\mathrm{bg} + I_\mathrm{pk} \exp\biggl[ -\frac{(\Delta x\cos\theta+\Delta y\sin\theta)^2}{2\sigma_x^2}\\
		-\frac{(-\Delta x\sin\theta+\Delta y\cos\theta)^2}{2\sigma_y^2} \biggr],
	\end{aligned}
\end{equation}
with the background flux level $I_\mathrm{bg}$, peak flux intensity $I_\mathrm{pk}$, $\Delta x = x-x_\mathrm{c}$, and $\Delta y = y-y_\mathrm{c}$. 
In Equation~(\ref{eqn:egauss}), $\theta$ is the position angle of the semi-major axis measured from the positive $x$-axis in the counterclockwise direction.

To determine $I_\mathrm{pk}$, semi-major and -minor axes, position angle, and their uncertainties from the best-fit model (and consequently the ellipticity of each object), we minimized the difference between the PSF-convolved model $M(x,y)$ and the postage-stamp galaxy image using the {\tt MPFIT} code \citep{mpfit}. We let the centroid $(x_{c}, y_{c})$ and the background level $I_\mathrm{bg}$ remain fixed to the SExtractor values ({\tt XWIN\_IMAGE}, {\tt YWIN\_IMAGE}) and {\tt BACKGROUND}, respectively, throughout the fitting. The ellipticity measurement errors were calculated by propagating the uncertainties of the best-fit parameters.

The ellipticity measurement obtained above needs to be calibrated. We performed WL image simulations and determined the global multiplicative correction factor of 1.22. Readers are referred to \citet{JT11PCA} and \citet{Jee13DLS} for details about the simulations. 
We applied the inverse-variance weighting scheme to estimate reduced shears, considering the dispersion of the ellipticity distribution and the measurement uncertainties as follows:
\begin{equation}
\label{eqn:g12calib}
g_{1(2)} = m_{1(2)}\frac{\sum_{i=1}^{N} e_{1(2)}\mu_{i}}{\sum_{i=1}^{N} \mu_{i}}.
\end{equation}
In the above equation, $\mu_{i}$ is the weight:
\begin{equation}
\label{eqn:eweight}
\mu_{i} = \frac{1}{\sigma_\mathrm{SN}^2 + (\delta e_{i})^2},
\end{equation}
$m_{1(2)}$ is the shear multiplicative factor, $\sigma_\mathrm{SN}$ is the shape noise (\mytilde0.25), and $\delta e_{i}$ is the measurement error per ellipticity component for the $i^{th}$ galaxy. 
We found that our additive bias (typically arising from PSF model bias) is negligible.

\subsubsection{Source Galaxy Selection \& Redshift Estimation} \label{subsubsec:srcgal}

As most lensing signals come from a faint population at high redshifts, ideally the redshifts from large spectroscopic or multi-wavelength photometric samples are needed to enable a clean separation of background sources. However, neither our current redshift nor photometric data provide the capability. Thus, we instead use the redshift-magnitude-color relation to select source galaxies and estimate their redshifts. Although less than ideal, this method has been used in numerous studies and proven to be practically useful for identifying substructures and quantifying their masses.

We defined our source population as the objects fainter and bluer than the A1240+A1237 red sequence.
The color--magnitude selection criteria are $23 < r < 27$ and $-1.0 < g-r < 0.8$. 
The bright-end is $\mytilde0.5$ mag fainter than the faintest spectroscopic cluster member 
to minimize the contamination by bright foreground galaxies and blue members while keeping the source density sufficiently high, 
whereas the faint-end is approximately the limiting mag in $r$. 
We also required sources to have well-defined shapes after their PSF effects are removed. Specifically, first, the elliptical Gaussian fitting must be successful (the {\tt MPFIT STATUS} parameter should be unity). Second, the ellipticity measurement error should be less than 0.3. Third, the (de-convolved) semi-major axis needs to be greater than 0.3 pixels. Finally, the ellipticity magnitude $(e_1^2+e_2^2)^{1/2}$ must be less than 0.9.
As illustrated in Figure~14 of \citet{Jee13DLS}, sources failing to meet these requirements usually form a characteristic pattern in the $e_1$ vs $e_2$ plot. 

A very bright star ($\mytilde9^{th}$ mag in $r$) is located near the cluster center ($\mytilde2\arcmin$ southwest 
of A1240N-BCG; see the panel (a) of Figure~\ref{fig:galmap}). Its reflection patterns and circular ghost rings make SExtractor detect numerous spurious objects around it, which also prevent us from identifying real galaxies in the region. Since the artifact is roughly axisymmetric in our coadd mainly because of the field rotation among visits, we modeled the feature as a function of radius from the star and subtracted the model from the coadd. Then, we ran SExtractor a second time on the star-subtracted image and measured source galaxy shapes that were hidden in the original coadd. We visually inspected every source galaxy in this region and a source is discarded if its shape is likely to be affected by subtraction residuals.

The objects in our final source catalog are shown as a green Hess diagram with square-root scaling in Figure~\ref{fig:cmd}. 
The mean number density is \mytilde43~arcmin$^{-2}$ in the central $27\arcmin\times27\arcmin$ region, where we perform our WL analysis. 

Equations~(\ref{eqn:kappa}) and (\ref{eqn:sigma_c}) show that the convergence $\kappa$ is a function of the angular diameter distance ratio $D_{ls}/D_{s}$. This dependence indicates that quantitative interpretation of the observed lensing signal relies on prior knowledge of the source redshift distribution. 
As mentioned above, we relied on the redshift-magnitude-color relation and utilized the external photometric redshift catalog as a reference to infer source redshifts in our cluster field. A potential concern in this approach is that the redshift-magitude-color relation in the reference field is significantly different from the one in our source population. \citet{Jee+2014elgordo} investigated this issue of cosmic variance on their mass determination using the Ultra Deep Field (UDF) and GOODS photometric redshift catalogs of \citet{udfcoe06} and \citet{goods2010}, respectively. Despite the large difference in field size, they found that the impact of the variance is small in mass estimation, quoting $\lesssim$4\%, which is sufficiently smaller than statistical uncertainties. 
In this study, we used the UVUDF catalog of \citet{uvudf2015}, who improved the result of \citet{udfcoe06} with the addition of WFC3 IR imaging data.

We account for the differences in depth and filter between the UVUDF and our data as follows. Photometric transformation of the ACS system to
the Subaru system was performed with the SED templates in \citet{bpzbenitez04} and the best-fit photo-$z$ values. We then applied our source selection criteria described above to this transformed UVUDF catalog.
We compared the magnitude distribution of this catalog to that of our sources and verified that the two results are consistent up to $r\lesssim25.5$. Beyond the threshold $r\gtrsim25.5$, our source catalog has gradually fewer objects per magnitude bin because of the shallower depth. Assuming that the color-magnitude-redshift relation in UVUDF is valid in our cluster field, we estimated the redshift distribution of our source galaxies after taking into account the difference in this magnitude distribution (completeness correction).

The lensing efficiency $\beta$ is defined as:
\begin{equation}
\beta = \biggl\langle\max\biggl[\frac{D_{ls}}{D_{s}}, 0\biggr]\biggr\rangle, \label{eqn:beta}
\end{equation}
where we set $\beta$ to zero for non-background sources, which do not contribute to the lensing signal. 
We obtained $\langle\beta\rangle=0.731$ and this corresponds to the effective redshift $z_\mathrm{eff}=0.868$. 
Since the lensing efficiency is nonlinear, the use of a single source plane at $z_\mathrm{eff}=0.868$ produces biased results. The first-order correction \citep{Seitz+1997,Hoekstra+2000} is:
\begin{equation}
g^{\prime} = \biggl[1 + \biggl(\frac{\langle\beta^{2}\rangle}{\langle\beta\rangle^{2}}-1\biggr)\kappa\biggr]g, \label{eqn:correctedg}
\end{equation}
\noindent
where $g^{\prime}$ is the uncorrected reduced shear.
We obtained $\langle\beta^{2}\rangle=0.579$, which reduces the observed shear $g^{\prime}$ by a factor of $(1+0.08\kappa)$. 

\subsubsection{Two-dimensional Weak-lensing Mass Reconstruction} \label{subsubsec:massmap}

The conversion of the measured ellipticities into the surface mass density ($\kappa$) was performed using the {\tt FIATMAP} code \citep{fiatmap97}, which implements the \citet{KS93} method in real space. 
To investigate the uncertainties in the mass density and peak centroids, 
we generated 1000 bootstrap realizations, from which
we derived the rms map of the convergence field.
The {\tt FIATMAP} convergence map was divided by the rms map to obtain the significance map.

\begin{figure*}
\begin{center}
\includegraphics[width=0.495\linewidth]{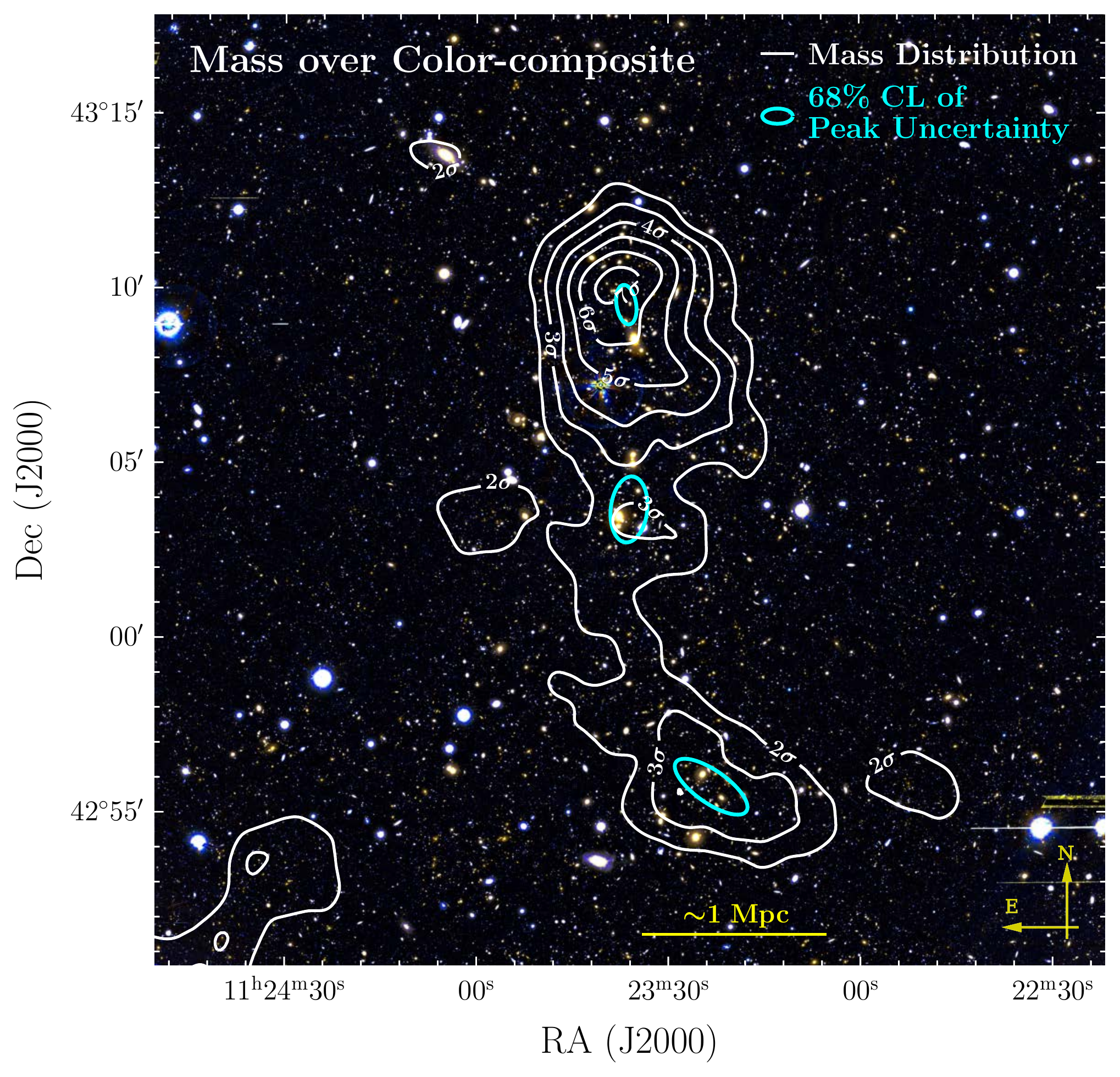}
\includegraphics[width=0.495\linewidth]{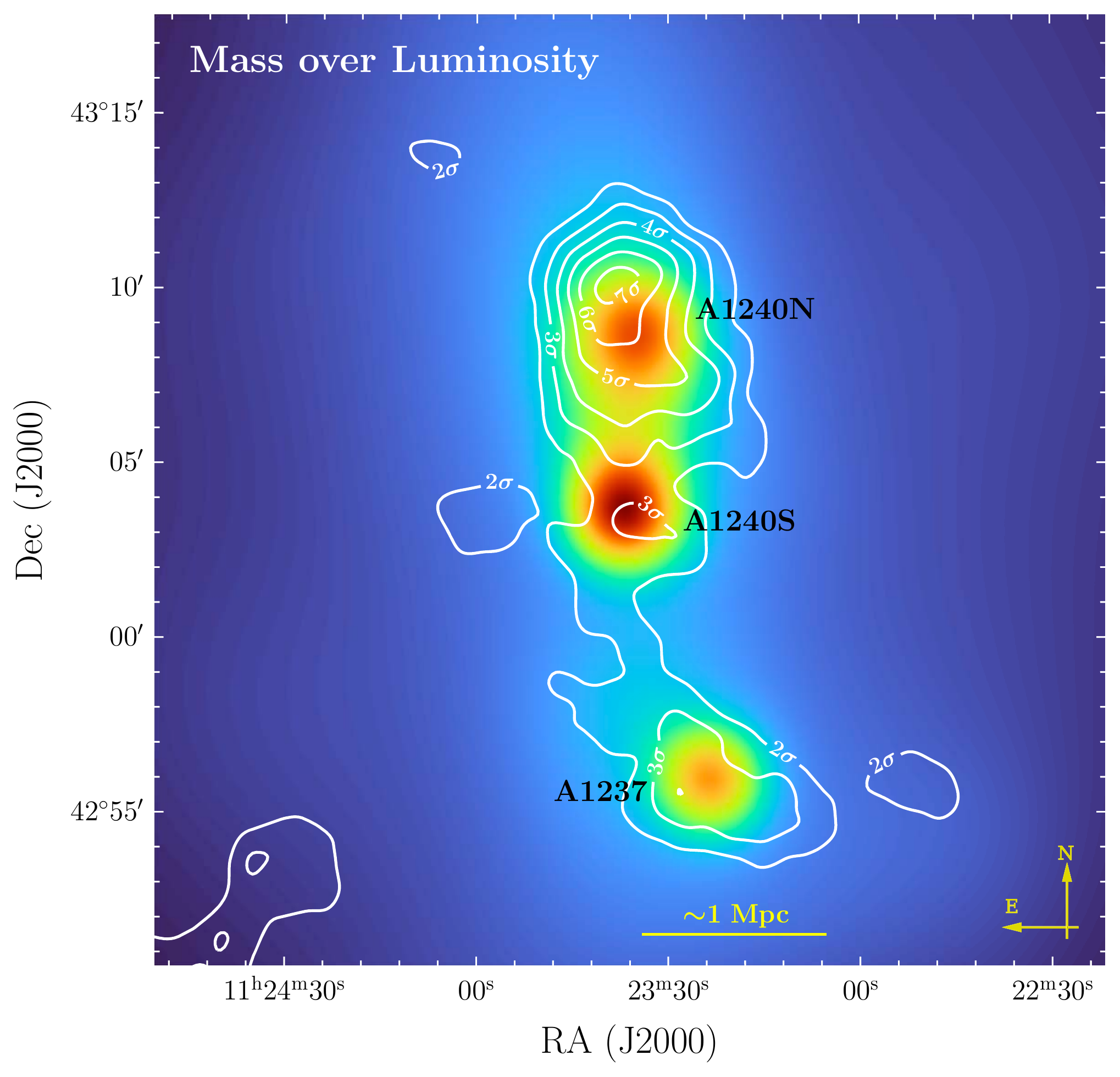}
\includegraphics[width=0.495\linewidth]{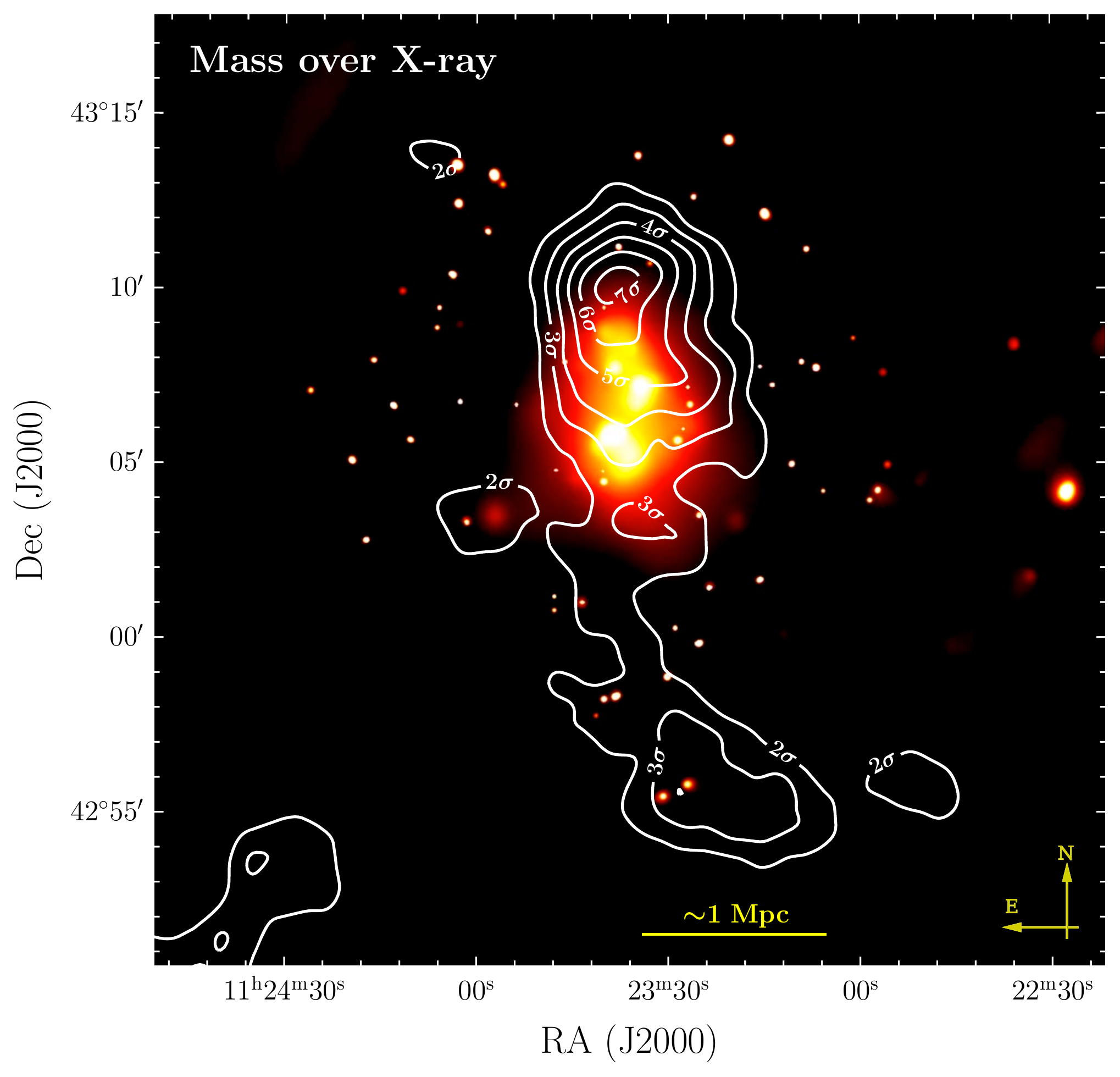}
\includegraphics[width=0.495\linewidth]{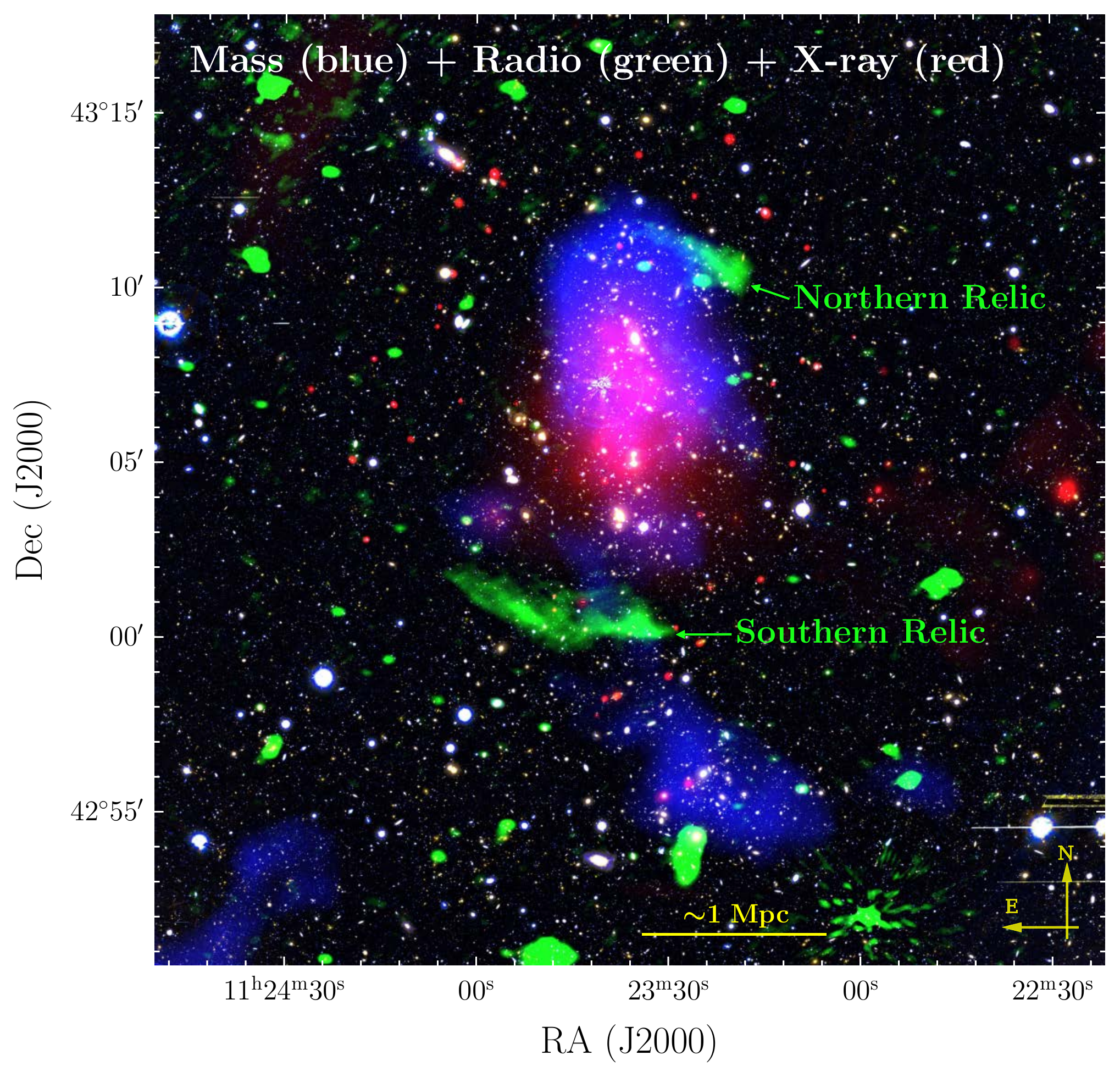}
\caption{Mass reconstruction of the A1240 field. We display our convergence ($\kappa$) significance map, which is obtained from division of the convergence map by the rms map. The rms is determined from 1000 bootstrapping runs. 
The lowest contour corresponds to the $2\sigma$ significance. 
We overlay the linearly-spaced significance contours on the Subaru color-composition image ({\it top left}), the galaxy luminosity-weighted density map ({\it top right}), and the 52~ks {\it Chandra} image ({\it bottom left}; only available for A1240).
In the top left panel, the 68\% covariance confidence limits of the mass centroid uncertainties are marked with cyan ellipses.
The composite image shown in the bottom right panel shows the mass, LOFAR 143~MHz radio \citep{Hoang+2018}, and {\it Chandra} X-ray data.
Our WL mass reconstruction detects three mass clumps associated with the three peaks in the optical luminosity. The A1240N, A1240S, and A1237 peaks are detected at the \mytilde$7.6\sigma$, \mytilde$3.3\sigma$, and \mytilde$4.1\sigma$ levels, respectively.
The distribution of the X-ray emission elongated in the N--S direction is approximately co-spatial with the optical luminosity. Compared with the mass, the extent of the X-ray emission is compact and mostly confined in-between the two mass clumps in A1240.
The northern and southern radio relics are located near the edges of A1240N
and A1240S defined by the optical luminosity.
\label{fig:massmap}}
\end{center}
\end{figure*}

In Figure~\ref{fig:massmap}, the convergence significance map is overlaid on the Subaru color-composite image (top left panel), the luminosity-weighted density map (top right panel), and
the 52~ks {\it Chandra} X-ray surface brightness map (bottom left panel).
Our mass reconstruction reveals that A1240 consists of northern and southern mass clumps, which are significant at the \mytilde$7.6\sigma$ and \mytilde$3.3\sigma$ levels, respectively, with a projected separation of \mytilde1.30~Mpc. The mass distribution of A1240 is elongated in the N--S direction, resembling the cluster galaxy luminosity distribution (A1240N and A1240S).
In addition, the third mass clump co-located with the luminosity peak A1237 is detected at the \mytilde$4.1\sigma$ level, \mytilde1.52~Mpc south of the A1240S mass clump. 
The $\mytilde68$\% peak distributions measured from the 1000 bootstrap-resampled mass maps show that all the three peaks are in good spatial agreement with the corresponding luminosity peaks.
Finally, the WL mass reconstruction detects a bridge connecting A1240 and A1237 at the $\geqslant2\sigma$ level. This filamentary extension in the intercluster region coincides with the overdensities of member galaxies. This is aligned with the \mytilde80~Mpc-long filamentary structure in the plane of sky (see \textsection\ref{subsubsec:filament}). 

The {\it Chandra} X-ray emission is also stretched in the same orientation as the A1240 mass and galaxies. 
In the bottom left panel of Figure~\ref{fig:massmap}, the X-ray surface brightness map shows a clumpy, elongated structure, 
from which it is somewhat difficult to quantify the exact number of the substructures. Nevertheless, it is clear that the gas extent is more compact (i.e., the gas filament is shorter than the baselines formed by the two mass/galaxy clumps, which is consistent with our expectation that A1240 is a post merger that happened in the N--S direction.

The bottom right panel of Figure~\ref{fig:massmap} presents 
the WL mass (blue), LOFAR 143~MHz intensity \citep[green;][]{Hoang+2018}, and {\it Chandra} X-ray surface brightness (red) maps together with the Subaru color-composite image.
The northern and southern radio relics, which bracket the two substructures of A1240 are stretched in the E--W direction, nearly perpendicular to the hypothesized merger axis.
\citet{Hoang+2018} found indications of possible discontinuities in X-ray surface brightness at the location of the relics, which is also expected if the radio relics indeed trace the merger shocks.

The southern relic is brighter and longer. Numerical simulations have shown that in general the kinetic energy flux, which is proportional to the relic brightness, associated with a less massive cluster in the binary collision is larger. Since our WL analysis suggests that A1240S is 2--3 times less massive (\textsection\ref{subsubsec:massfit}), this disparity in relic size is also consistent with our expectation.

As for A1237, since its redshift is similar to that of A1240, it is an important question whether it has already interacted with A1240. \citet{Hoang+2018} detected a tailed radio galaxy (we denoted it with A1237BCG-S in the panel (a) of Figure~\ref{fig:galmap}). Since it has an extension toward south, the authors suggest that A1237BCG-S might be moving through the local ICM toward the A1237 center. However, this does not help us to infer the bulk motion of A1237. 
Since A1237 is not fully covered by the {\it Chandra} data, 
currently the question still remains unresolved.
Nevertheless, from the symmetric X-ray and radio features 
and the large separation from A1240, 
our conjecture is that A1237 is not likely to be involved in the recent A1240 N--S merger responsible for the two distinct radio relics and the characteristic X-ray morphology. 

\subsubsection{Mass Estimation} \label{subsubsec:massfit}

As we detected two mass components in A1240 and a separate clump in A1237, we cannot treat the mass distribution of the A1240 field as a single halo. 
To quantify the mass of each subcluster, 
we performed Markov Chain Monte Carlo (MCMC) analysis by simultaneously fitting three halos.
For each halo, we assumed a Navarro-Frenk-White \citep[NFW;][]{NFW1996, NFW1997} profile with the mass--concentration relation of \citet{duffy08} and fixed their centroids at the luminosity centers.
We excluded source galaxies at the core of each halo located within $r_\mathrm{min}=10\arcsec$.
During the likelihood sampling, we imposed a flat prior on mass with the interval $10^{13}$\Msun\ $<M_{200}<10^{15}$\Msun.

\begin{figure}
\centering
\includegraphics[width=1.0\linewidth]{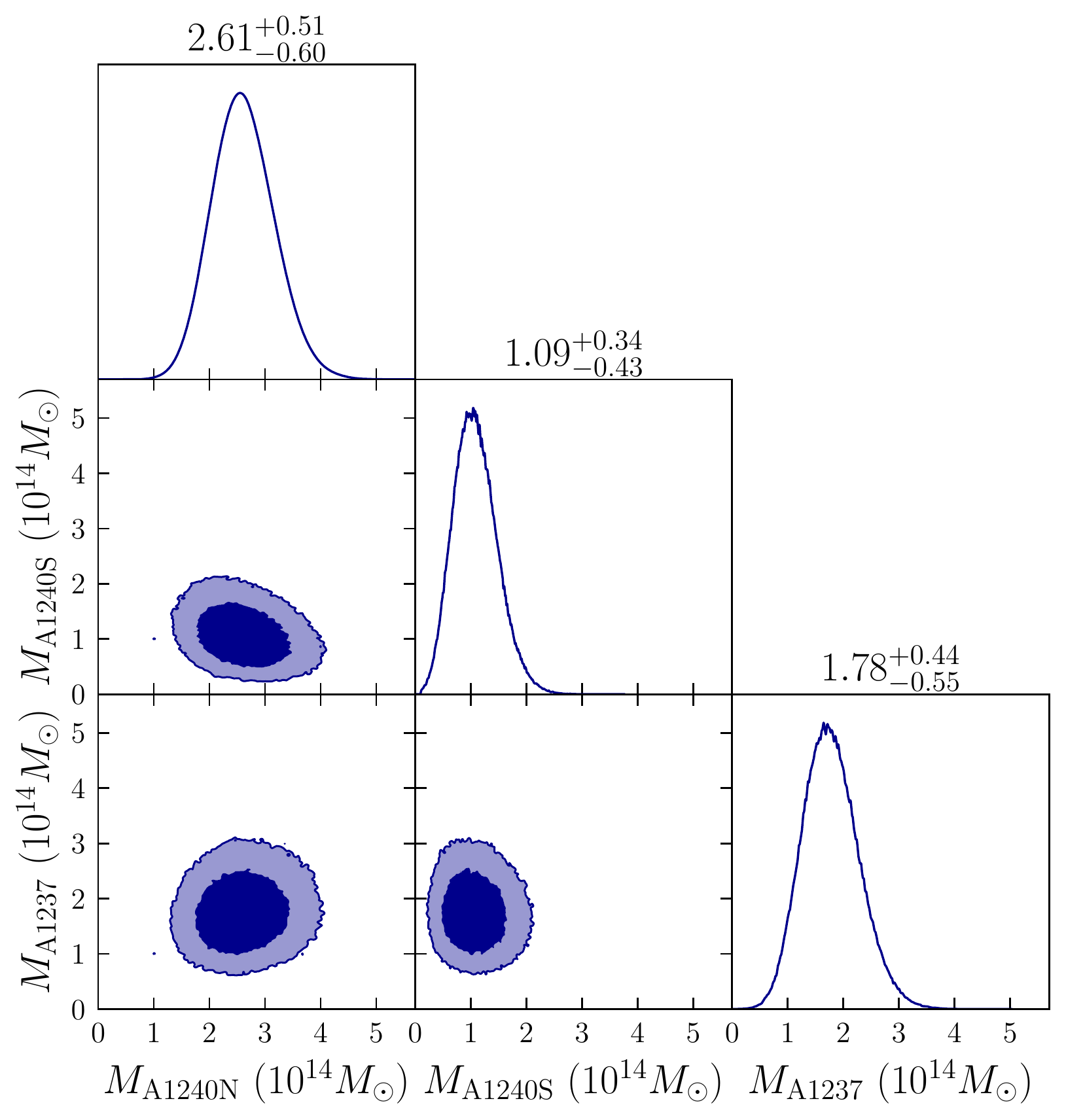} 
\caption{Posterior distributions for the masses of A1240 northern (A1240N) and southern (A1240S) subclusters and A1237 from MCMC analysis. 
While fitting three halos with NFW profiles simultaneously, we assumed the \citet{duffy08} mass--concentration relation based on the flat $\Lambda$CDM cosmology. 
The filled contours show 68\% and 95\% confidence limits for marginalized constraints, evaluated from the MCMC chains using the {\tt GetDist} code \citep{getdist}.
}
\label{fig:massmcmc}
\end{figure}

Figure~\ref{fig:massmcmc} shows the MCMC results.
The $M_{200}$ values of the northern and southern halos of A1240 are $M_\mathrm{A1240N}=2.61_{-0.60}^{+0.51}\times10^{14}$\Msun\ and $M_\mathrm{A1240S}=1.09_{-0.43}^{+0.34}\times10^{14}$\Msun, respectively, which show that A1240
is a merger with a mass ratio of \mytilde2:1.
We also estimated A1237’s mass to be $M_{A1237}=1.78_{-0.55}^{+0.44}\times10^{14}$\Msun. 
The mass estimation results are summarized in Table~\ref{tab:subclusters} and utilized for our merger scenario reconstruction in \textsection\ref{subsec:scenario}.

To estimate the A1240 total system mass, we adopted the approach described in \citet{Jee+2014elgordo}. 
We assumed that the global center of A1240 is the geometric mean of the A1240N and A1240S luminosity peaks. 
Then, the 3D mass distribution of the entire system is modeled as a superposition of two NFW profiles.
For each MCMC sample, we numerically determined the radius from the global center that encloses $M_{200}$.
The resulting A1240 system mass is $M_{200}=(4.37\pm0.77)\times10^{14}$\Msun\ with $r_{200}=1.47\pm0.09$~Mpc.
Since many previous studies quoted the A1240 mass using $M_{500}$, it is also useful to compute $M_{500}$ with our WL result for fair comparison. The result is $M_{500}=(2.83\pm0.50)\times10^{14}$\Msun. However, since this $M_{500}$ value is reached at $r\sim0.9$~Mpc, which is less than the distance between A1240N and A1240S, we believe that this $M_{500}$ value does not adequately represent the global property of A1240.

\section{Discussion} \label{sec:discussion}

Both identification and quantification of the A1240 substructures based on our multi-wavelength studies are critical in our reconstruction of the merging scenario.
Here, we compare our key measurements with previous studies and present our merging scenario reconstruction.

\subsection{Comparison with Previous Studies}

\subsubsection{Kinematics Comparison} \label{subsubsec:comp-kin}

The kinematical properties of the two subclusters in A1240 were studied by \citetalias{Barrena+2009} and \citetalias{Golovich+2019b}. 
Using BCGs as tracers of merging subclumps, \citetalias{Barrena+2009} reported that the LOS rest-frame velocity difference between the northern and southern subclumps is $\mytilde$390~\kms, which is consistent with the value estimated by \citetalias{Golovich+2019b} ($394\pm117$~\kms). 
However, we obtained a relatively small difference ($83\pm185$~\kms) from the nearby member galaxies around the BCGs (\textsection\ref{subsubsec:galdens}).
A small LOS velocity difference indicates that the merger might be happening close to the plane of the sky and/or that A1240N and A1240S might be observed near their apocenters \citep[e.g.,][]{Dawson+2015MC2_CIZAJ2242,Golovich+2016MC2_MACSJ1149,vanWeeren+2017A3411-3412}.
The \mytilde$2\sigma$ discrepancy is mainly due to the subcluster membership assignment.
We only considered the objects in the dense core regions within \mytilde0.5~Mpc from the center of each luminosity density peak whereas previous studies covered an area up to
9 times larger.
Since the two subclusters are close ($\mytilde1$~Mpc) while the merger is expected to cause significant outflows of the member galaxies, we argue that our conservative choice is safer in the membership identification.

Looking only at the subcluster cores, we find that our result is in agreement with \citetalias{Barrena+2009}, who also found that the inner regions of their subclusters seem to have similar velocities (see their Figures 9 and 10).
Their velocity profile for the member galaxies of A1240N gradually increases outward, regardless of the choice of the subcluster center. 
\citetalias{Barrena+2009} suggested that perhaps a few galaxies at higher redshift in the outer region of the northern cluster leads to the relatively high mean redshift of A1240N (thus, a large LOS velocity difference).
The projected locations of these galaxies correspond to the candidate cluster GMBCG170.89---the northern yellow star inside the green dash-dotted circle in the right panel of Figure~\ref{fig:filament}---at a slightly higher redshift than that of A1240. 
GMBCG170.89 appears to be part of the N--S filamentary structure. 
The presence of these galaxies could be interpreted as either a group of galaxies possibly accreting onto A1240 through the filament or a plume of outflowing galaxies at large clustercentric distances due to the recent merging event between A1240N and A1240S (as suggested for A1758 by, e.g., \citealt{A1758N_Boschin+2012,A1758_Schellenberger+2019}; A3266 by \citealt{A3266_Quintana+1996,A3266_Flores+2000}; and Cl0024+1654 by \citealt{Cl0024+1654_Czoske+2002}).

\citetalias{Barrena+2009} provided the first measurements of the LOS velocity dispersions for the two individual subclusters in A1240 using 32 and 27 member redshifts, respectively. They quoted $709_{-83}^{+88}$~\kms\ ($991_{-99}^{+149}$~\kms) for A1240N (A1240S). 
With a factor of \mytilde2 more redshifts, \citetalias{Golovich+2019b} reported that both subclusters have similar velocity dispersions ($706\pm52$~\kms\ and $727\pm68$~\kms\ for A1240N and A1240S, respectively). 
Considering only 45 (51) members in the core region of A1240N (A1240S), we obtained $909\pm116$~\kms\ ($841\pm109$~\kms). 
These velocity dispersions cannot be used as reliable mass proxies because the merger must have made the system depart from the dynamical equilibrium; when we convert our WL measurement to a velocity dispersion under the assumption of SIS, we obtain $540_{-55}^{+50}$~\kms\ ($464_{-63}^{+55}$~\kms) for A1240N (A1240S), which is significantly lower than the spectroscopic value.

\subsubsection{Mass Comparison} \label{subsubsec:comp-mass}

The quoted uncertainties of our mass estimates include the statistical contributions from the intrinsic shape noise and ellipticity measurement errors. There are additional systematic uncertainties to keep in mind as one interprets our results. As noted in \citet{Jee+2014elgordo}, scatter in the mass--concentration relation, halo triaxiality, departure from the NFW assumption, and uncorrelated large-scale structures along the LOS can increase the total mass uncertainty up to \mytilde30\%.

\begin{figure}
\centering
\includegraphics[width=1.0\linewidth]{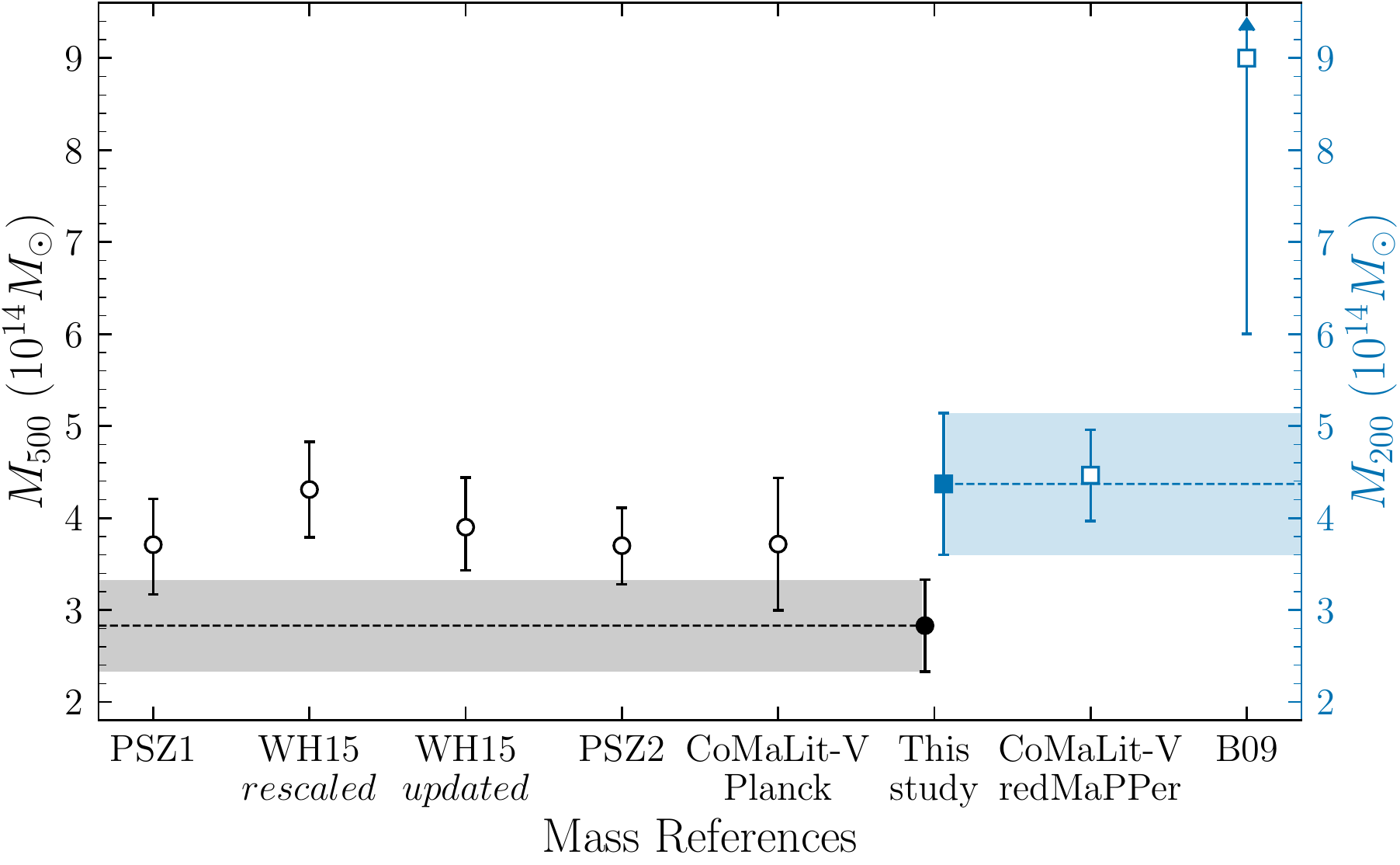} 
\caption{A1240 WL mass comparison with other estimates.
Black circles and blue squares are $M_{500}$ (read the $y$-axis on the left) and $M_{200}$ (read the $y$-axis on the right), respectively, in units of $10^{14}$\Msun. 
Filled symbols (and dashed lines) and shaded regions indicate our mass estimates and the corresponding $1\sigma$ uncertainties, respectively, from the current WL study. 
Overall, our results are broadly consistent with previous studies in both $M_{500}$ and $M_{200}$. Note that for the WL evaluation of $M_{500}$ our choice of the cluster center (i.e., geometric center between A1240N and A1240S) may cause underestimation because the mass density is low there. See text for abbreviated names of mass references.}
\label{fig:masscomp}
\end{figure}

Figure~\ref{fig:masscomp} summarizes the comparison of the total mass estimates $M_{500}$ and $M_{200}$ of A1240 in the literature with our WL results. 
The first mass estimation of A1240 is presented in \citetalias{Barrena+2009}, based on velocity dispersion measurements. 
Under the assumption of the dynamical equilibrium, they quoted 
$M_\mathrm{vir}=(0.9\textrm{--}1.9)\times10^{15}$\Msun\ within 
$r_\mathrm{vir}=1.9\textrm{--}2.4$~Mpc for the entire A1240 system. 
The \citetalias{Barrena+2009} result is significantly larger than our WL estimate of the total mass $(4.37\pm0.77)\times10^{14}$\Msun\ (\textsection\ref{subsubsec:massfit}). 
Possible causes of this large discrepancy were discussed by \citet{mass_Pinkney+1996} and \citet{mass_Takizawa+2010}, who investigated the dependence of the mass prediction through virial theorem on the merger status using numerical simulations of head-on binary mergers. 
They concluded that the cluster mass estimation through the virial theorem would severely be affected by the merger. 
The bias is caused not only by the departure from the equilibrium, but also by the viewing angle of the merger. 
For example, according to \cite{mass_Pinkney+1996}, when a 3:1 mass ratio merger is viewed within \mytilde$60\degr$ of the LOS, the virial mass would be overestimated by up to a factor of two.
Similarly, \cite{mass_Takizawa+2010} showed that when a merger is observed along the LOS, 
a factor of two or more overestimation could occur. 

Using the richness--mass ($\lambda$--$M_{200}$) relation, \citet[][abbreviated ``CoMaLit-V redMaPPer'' in Figure~\ref{fig:masscomp}]{Sereno+17CoMaLit5} performed mass-forecasting of A1240. 
They used the subset of ``Literature Catalogues of weak Lensing Clusters of galaxies'' \citep{Sereno+15CoMaLit3} 
compiled from the WL masses of galaxy clusters in the literature that have counterparts in the SDSS redMaPPer catalog 
as a calibration sample to obtain the median scaling relation.
The resulting mass estimate $M_{200}=(4.47\pm0.50)\times10^{14}$\Msun\ is in good
agreement with our WL result. 

\citet[]{PSZ1+2015,PSZ2+2016}, who referred to A1240 as PSZ1 G165.41+66.17 and PSZ2 G165.46+66.15, respectively, estimated $M_{500}$ to be $M_\mathrm{PSZ1}=3.71_{-0.54}^{+0.50}\times10^{14}$\Msun\ and $M_\mathrm{PSZ2}=3.70_{-0.42}^{+0.41}\times10^{14}$\Msun\ based on the SZ signal.
\citet[][abbreviated ``WH15'' in Figure~\ref{fig:masscomp}]{WH15} quoted a mass of 
$M_{500,\textit{rescaled}}=(4.31\pm0.52)\times10^{14}$\Msun\ with their calibrated optical mass proxy; their updated richness $R_{L*,500}=65.45$ of A1240 can be converted into the mass of $M_{500,\textit{updated}}=3.90_{-0.47}^{+0.54}\times10^{14}$\Msun.
The mass estimate of A1240 in \citet[][abbreviated ``CoMaLit-V Planck'' in Figure~\ref{fig:masscomp}]{Sereno+17CoMaLit5} is $M_{500}=(3.72\pm0.72)\times10^{14}$\Msun.
Our WL estimate $M_{500}=(2.83\pm0.50)\times10^{14}$\Msun\ is marginally consistent with the above, although we believe that our choice of the cluster center biases the measurement lower (see the caveat mentioned in \textsection\ref{subsubsec:massfit}).

For the northern and southern subclusters individually, \citetalias{Barrena+2009} obtained $M_\mathrm{vir}=(5\pm2)\times10^{14}$\Msun\ and $(14\pm5)\times10^{14}$\Msun\, respectively, from their LOS velocity dispersion measurements. 
According to \citetalias{Barrena+2009}, A1240S is about three times more massive.
However, when \citet{Wittman+2018} used the \citetalias{Golovich+2019b} values of the subcluster velocity dispersion, they obtained similar masses [$M_\mathrm{A1240N}=(4.19\pm0.99)\times10^{14}$\Msun\ and $M_\mathrm{A1240S}=(4.58\pm1.40)\times10^{14}$\Msun].
Our WL mass estimation reveals that the mass ratio is about 2:1 ($M^\mathrm{A1240N}_{200}=2.61_{-0.60}^{+0.51}\times10^{14}$\Msun\ and $M^\mathrm{A1240S}_{200}=1.09_{-0.43}^{+0.34}\times10^{14}$\Msun).
Because the southern radio relic is significantly larger than the northern one, this 2:1 mass ratio better agrees with our understanding that in general the kinetic energy flux associated with the less massive system is greater. 

\subsection{Merging Scenario} \label{subsec:scenario}

Cluster mergers happen in a timescale of a few Gyrs while we only witness a single snapshot. To utilize a cluster merger as a useful astrophysical laboratory, it is essential to constrain the stage of the merger, which is often represented by parameters such as time-since-collision (TSC), collision velocity, viewing angle, impact parameter, apocenter position, etc.
Although one of our ultimate goals in \mcc\ is to reproduce many observed features with sophisticated high-resolution numerical simulations \citep[e.g.,][]{LeeW+2020}, finding the optimal combination of simulation setups is still a very challenging task limited by severe degeneracies among different parameters and huge computational time. Here, we employ the Monte-Carlo Merger Analysis Code \citep[{\tt MCMAC};][]{Dawson2013,mcmac2014}, which enables fast prediction on the 3D configurations of a cluster merger, based on observed features and Monte-Carlo simulations.

The {\tt MCMAC} method employs several assumptions in its analysis of the merger dynamics. They include mass and energy conservations and zero angular momentum throughout the merger history. Thus, each subcluster maintains a constant mass (i.e., input mass from observation). Furthermore, subclusters interact only through gravity. Since the model treats the merger dynamics as isolated, collisionless, binary system, the effects of the surrounding large-scale structure, dynamical fraction, and tidal stripping of DM and gas are ignored. 
Despite these caveats, in general, the {\tt MCMAC} method achieves a good (\mytilde10\%) accuracy in estimating some important dynamical parameters when compared with hydrodynamic $N$-body simulation results \citep{Dawson2013}. A recent study by \citet{KimJ+2021_elgordo}, however, showed that the {\tt MCMAC} code produces severely biased results when applied to the ``El Gordo" merging cluster, where the dynamical friction becomes no longer negligible because of the extreme masses of individual subclusters. Fortunately, the moderate mass of A1240 makes the omission of the dynamical friction in {\tt MCMAC} much less critical in our merging scenario analysis.

The key input parameters to {\tt MCMAC} are the redshifts and masses of A1240N and A1240S, projected separation between the two, and their measurement uncertainties. The {\tt MCMAC} output includes collision velocity, viewing angle, and TSC for each combination of the randomized (within their measurement uncertainties) masses, redshifts, and projected separation. The degeneracies among the output parameters can be reduced by additional constraints such as the polarization fraction and positions of radio relics.
\cite{Hoang+2018} measured the mean fractional polarization of A1240N and A1240S to be $32\pm4$\% and $17\pm4$\%, respectively, from the VLA 2--4~GHz data. 
We translated these measurements into the viewing angle prior $\alpha<51\degr$, where $\alpha$ is the angle between the merger axis and the plane of the sky.
In order to utilize the observed relic positions to further constrain the merger state, we need to model the propagation of the relic. Numerical simulations have shown that the merger shock propagation velocity $v_\mathrm{relic}$ (with respect to the center of mass; CM) is 
related to the associated subcluster collision velocity $v^\mathrm{CM}_\mathrm{col}$ (with respect to the CM) at pericenter via $v_\mathrm{relic}\sim \eta v^\mathrm{CM}_\mathrm{col}$, where $\eta$ is close to unity \citep[e.g.,][]{Springel_Farrar2007,Paul+2011}, although the exact value is unknown. We adopt $\eta=0.9$ as the fiducial value in this paper, following \citet{Ng+2015}.

\begin{figure}
\centering
\includegraphics[width=1\linewidth]{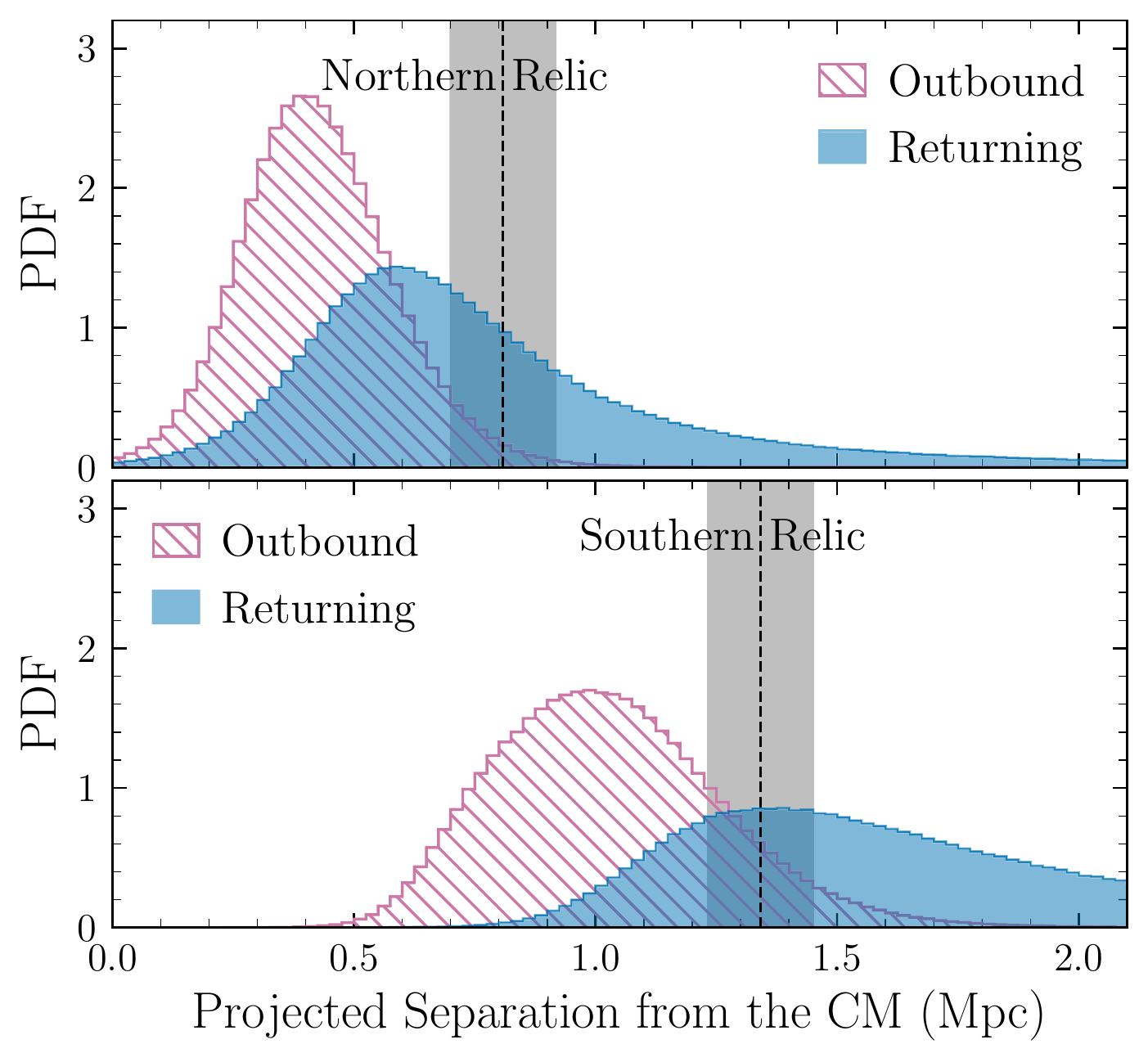}
\caption{Probability density function of the projected separation of the merger shock from the center of mass (CM). 
The vertical dashed lines and grey-shaded regions indicate the observed locations and positional uncertainties of radio relics, respectively. The relic positional uncertainties are dominated by the CM positional uncertainties, which are in turn affected by the mass uncertainties.
The northern relic favors a returning scenario with $p_{r}/p_{o}\sim4.9$, where $p_{r}$ and $p_{o}$ are the probabilities of the returning and outbound cases, respectively, integrated over the relic position prior interval ({\it top}). 
On the other hand, the difference is small ($p_{r}/p_{o}\sim1.3$) with the southern relic ({\it bottom}).
Combining the results from both panels (counting the Monte-Carlo samples satisfying both constraints), we obtain $p_{r}/p_{o}\sim37.5$, which shows that the returning scenario is strongly preferred.
}\label{fig:sproj}
\end{figure}

\begin{figure*}
\centering
\includegraphics[width=1\linewidth]{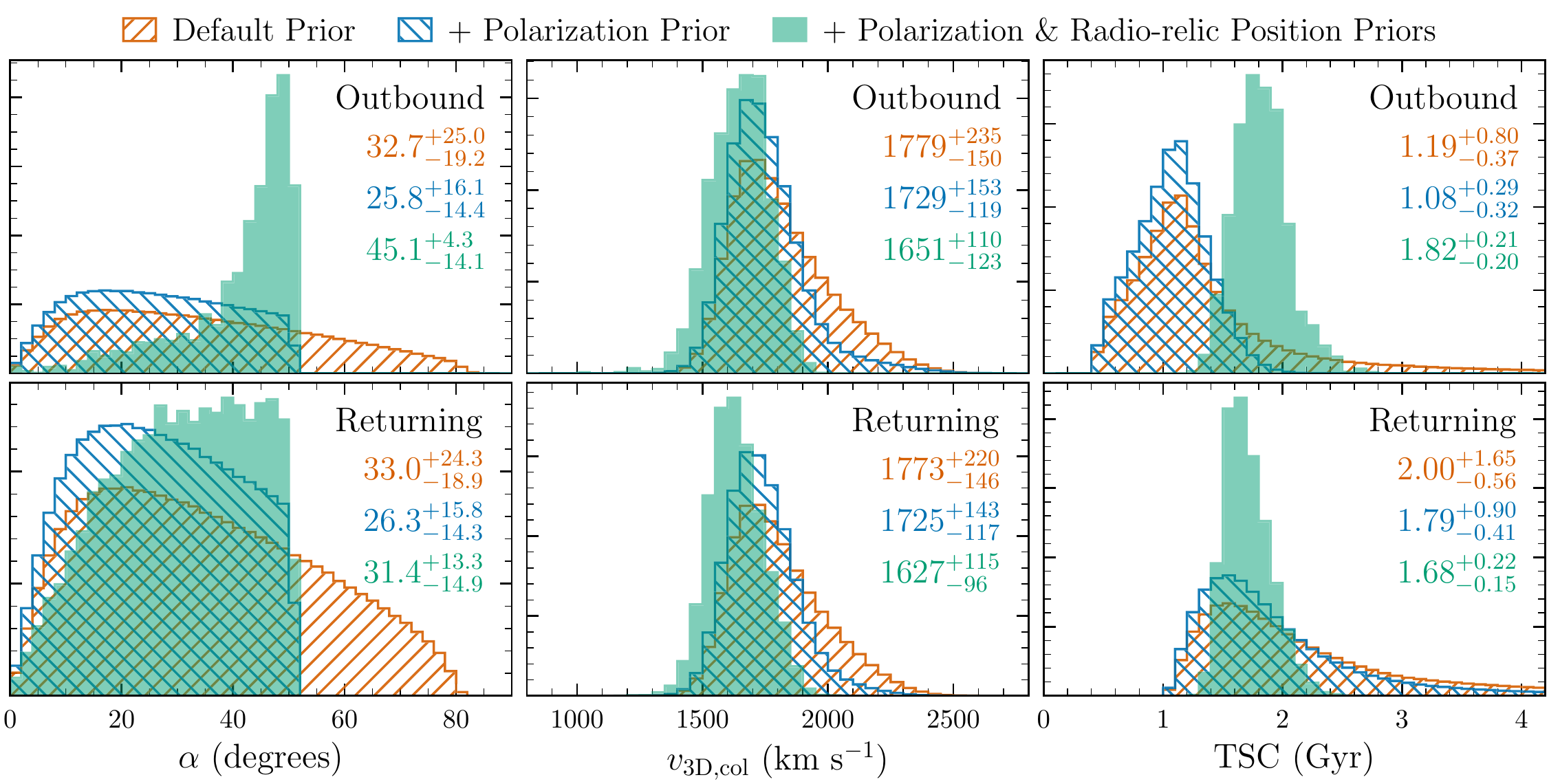}
\caption{Marginalized PDFs of the A1240 merger parameters. 
With different choices of priors, we display the results for $\alpha$, the merger axis angle with respect to the plane of the sky, $v_{\rm 3D,col}$, the three-dimensional relative velocity at the pericenter, and TSC for the outbound ({\it top}) and returning ({\it bottom}) cases. 
}\label{fig:mcmac}
\end{figure*}

Figure~\ref{fig:sproj} shows the probability density functions (PDF) of the radio relic positions obtained from the {\tt MCMAC} result. For the northern relic, the observed relic position favors a returning scenario ($p_{r}/p_{o}\sim4.9$) whereas the difference is small ($p_{r}/p_{o}\sim1.3$) with the southern relic.
When we combine the constraints from both relics (by counting the {\tt MCMAC} samples that satisfy both relic positions), a returning scenario is strongly favored with $p_{r}/p_{o}\sim37.5$.

We compare the viewing angle, collision velocity, and TSC for different priors in Figure~\ref{fig:mcmac}.
For the returning case, when we employ both polarization and radio-relic position priors, the viewing angle, collision velocity, and TSC are constrained to be $\alpha=31.4_{-14.9}^{+13.3}$~deg, $v_\mathrm{3D,col}=1627_{-96}^{+115}$\kms, and $\mathrm{TSC}=1.68_{-0.15}^{+0.22}$~Gyr, respectively. Although we find that consistent results are obtained for different choices of priors, the differences in uncertainty are significant. For example, when no prior is used, the uncertainty of TSC increases by a factor of $\mytilde6$ compared to the case when both polarization and radio-relic position priors are used.
For the outbound case, which is not favored by the radio relic position, the collision velocities are similar to the results in the returning case. Interestingly, in this scenario the most probable viewing angle is somewhat large ($\alpha\gtrsim45\degr$) with the polarization and radio relic priors. We believe that this happens because the projected shock propagation speed must be substantially lowered in order to match the observed relic positions; in this case the resulting TSC should also be made even longer than the value in the returning case. 

Note that our merger scenario is different from that of
\citet{Wittman2019}, who investigated the dynamical properties of A1240 with simulated analogs and reported that an outbound case is more probable (84\% vs 16\%). Although the method is different from ours, the discrepancy arises mostly from the fact that \citet{Wittman2019} adopted larger masses 
($\mytilde4.2\times10^{14}M_{\sun}$ and $\mytilde4.6\times10^{14}M_{\sun}$ for A1240N and A1240S, respectively) inferred from the subcluster velocity dispersions and also did not use the radio relic position constraint. For example, the 68\% collision velocity interval
(1979--2466~\kms) estimated by \citet{Wittman2019} roughly corresponds to the escape velocity of the A1240 system computed with the current WL masses. 

\section{Conclusions} \label{sec:summary}

We have studied the double radio relic cluster A1240 and its companion cluster A1237 with multi-wavelength data. Our findings are summarized as follows.
\begin{itemize}
\item Our MMT/Hectospec observation increases the number of the spectroscopic members to 432, which is a factor of $\mytilde3$ larger than the previous value. Combining the result with our Subaru-based photometric data, we were able to identify the cluster substructures in unprecedented detail. 
\item Our Subaru-based weak lensing analysis detects three significant mass clumps with masses of $M_\mathrm{A1240N}=2.61_{-0.60}^{+0.51}\times10^{14}$\Msun, $M_\mathrm{A1240S}=1.09_{-0.43}^{+0.34}\times10^{14}$\Msun, and $M_{A1237}=1.78_{-0.55}^{+0.44}\times10^{14}$\Msun\ forming a $\mytilde4$~Mpc filamentary structure in the north-south orientation. These three mass peaks are in good spatial agreement with the cluster galaxy distributions obtained with our spectroscopic and photometric member catalog. 
\item The northern (A1240N) and middle (A1240S) mass clumps separated by $\mytilde1$~Mpc are associated with A1240 and co-located with the X-ray emission detected with the {\it Chandra} data. The double radio relics bracket the edges of these two subclusters. 
\item Our Monte-Carlo analysis shows that the current positions of the double radio relics strongly constrain the merger phase, favoring a returning scenario with $\mathrm{TSC}\sim1.7$~Gyr. 
\item With the SDSS DR16 data analysis, we find that A1240 is embedded in the much larger-scale (\mytilde80~Mpc) filamentary structure whose orientation is in remarkable agreement with the hypothesized merger axis of A1240. 
\end{itemize}

Radio relic clusters are receiving a growing attention as powerful astrophysical laboratories, which enable useful experiments that are impossible on the ground to probe properties of dark matter, mechanisms of cosmic ray particle acceleration, evolution of plasma turbulence, etc. 
The rarest of the relic classes are symmetric double relics, which are equidistant from the cluster center with comparable surface brightness and provide two redundant probes of the same merger. To date, about six objects that meet our criteria have been reported, and A1240 is one of the cleanest systems that exhibit a distinct bimodal mass structure.
One important future scientific objectives of our \mcc\ project is to follow up these priority targets with high-fidelity numerical simulations and address the aforementioned scientific questions. Our multi-wavelength study of A1240 presented here serves as a principal stepping stone to this goal.


\indent

\indent

\noindent
This work was supported by K-GMT Science Program (PID: MMT-2019A-004) of Korea Astronomy and Space Science Institute (KASI). 
Observations reported here were obtained at the MMT Observatory, a joint facility of the Smithsonian Institution and the University of Arizona.
This work is based in part on data collected at Subaru Telescope and obtained from the SMOKA, which is operated by the Astronomy Data Center, National Astronomical Observatory of Japan.
HC and KF thank Ho Seong Hwang for providing guidance on the MMT/Hectospec data reduction.
HC acknowledges support from the Brain Korea 21 FOUR Program. 
MJJ acknowledges support for the current research from the National Research Foundation (NRF) of Korea under the programs 2017R1A2B2004644 and 2020R1A4A2002885.

\vspace{5mm}
\facilities{Subaru(Suprime-Cam), MMT(Hectospec), CXO, LOFAR}

\software{
Astropy \citep{astropy:2013, astropy:2018}, CIAO \citep{ciao}, {\tt emcee} \citep{emcee2013}, {\tt FIATMAP} \citep{fiatmap97}, {\tt GetDist} \citep{getdist}, HSRED (\url{https://github.com/MMTObservatory/hsred}), IRAF/RVSAO \citep{Kurtz1998}, Matplotlib \citep{matplotlib}, {\tt MPFIT} \citep{mpfit}, {\tt MCMAC} \citep{mcmac2014}, NumPy \citep{numpy:2020}, SCAMP \citep{scamp}, SciPy \citep{scipy:2020}, SDFRED2 \citep{sdfred2}, SExtractor \citep{sextractor}, SWarp \citep{swarp}
}

\bibliography{ms-a1240}
\bibliographystyle{aasjournal}

\end{document}